\documentclass[seceq]{ptptex}

\usepackage{graphicx}
\usepackage{wrapft}

\def\lesim{\m@thcombine<\sim}
\def\gesim{\m@thcombine>\sim}
\def\lessgtr{\m@thcombine<>}
\def\gtrless{\m@thcombine><}

\newcommand{\Se}{${}^{68}$Se}

\newcommand{\bra}[1]{\left\langle #1 \right|}
\newcommand{\ket}[1]{\left| #1 \right\rangle}
\newcommand{\Hhat}{\hat{H}}
\newcommand{\HhatMq}{\hat{H}_M(q)}
\newcommand{\hhat}{\hat{h}}
\newcommand{\hhatq}{\hat{h}(q)}
\newcommand{\hhatMq}{\hat{h}_M(q)}
\newcommand{\Fhat}{\hat{F}}

\newcommand{\Fhatp}{\hat{F}^{(+)}}
\newcommand{\Fhatm}{\hat{F}^{(-)}}
\newcommand{\Hc}{{\cal H}}
\newcommand{\Nhat}{\hat{N}}

\newcommand{\Qhat}{\hat{Q}}
\newcommand{\Rhat}{\hat{R}}
\newcommand{\Phat}{\hat{P}}

\newcommand{\phiqppn}{\phi(q,p,\varphi,N)}
\newcommand{\phiqpn}{\phi(q,p,N)}

\newcommand{\phiq}{\phi(q)}

\newcommand{\del}{\partial}

\newcommand{\beq}{\begin{equation}}
\newcommand{\beqa}{\begin{eqnarray}}
\newcommand{\eeq}{\end{equation}}
\newcommand{\eeqa}{\end{eqnarray}}

\def\vect#1{\mbox{\boldmath $#1$}}

\newcommand{\HPi}{\frac{G_{\tau}}{2} \Big( A_{\tau}^{\dag}A_{\tau}+A_{\tau}A_{\tau}^{\dag} \Big)}
\newcommand{\HQQi}{\frac{\chi}{2} \sum_{K=-2}^{2}D_{2K}^{\dag}D_{2K}}

\newcommand{\HS}{\sum_{k}\epsilon_{k}c_{k}^{\dag}c_{k}}

\newcommand{\PAIR}{\sum_{k \in \tau}d_{k}^{\dag}d_{{\bar k}}^{\dag}}

\newcommand{\nmu}{\bar {\mu}}
\newcommand{\nk}{\bar {k}}
\newcommand{\nnu}{\bar {\nu}}
\newcommand{\nl}{\bar {l}}

\def\expect#1{\langle #1 \rangle }
\def\matel#1#2#3{\langle #1|#2|#3\rangle}

\newcommand{\amndag}{\vect A_{\mu \nnu}^{\dag}}
\newcommand{\amn}{\vect A_{\mu \nnu}}
\newcommand{\ammdag}{\vect A_{\mu \nmu}^{\dag}}
\newcommand{\amm}{\vect A_{\mu \nmu}}

\newcommand{\xmndag}{\vect X_{\mu \nnu}^{(i)\dag}}
\newcommand{\xmn}{\vect X_{\mu \nnu}^{(i)}}

\newcommand{\fq}{f_{Q,s}^{(-)}}

\newcommand{\fpr}{f_{PR,s}^{(+)}}

\newcommand{\ks}{\kappa_{s}}

\newcommand{\FAspni}{F_{s}^{(+)}}
\newcommand{\FAsmni}{F_{s}^{(-)}}
\newcommand{\RAspni}{R_{s}^{(+)}}
\newcommand{\RAsmni}{R_{s}^{(-)}}

\newcommand{\FAsppni}{F_{s'}^{(+)}}
\newcommand{\FAsmpni}{F_{s'}^{(-)}}

\newcommand{\RAsmpni}{R_{s'}^{(-)}}

\newcommand{\SQQ}{S_{s s'}^{Q,Q}}
\newcommand{\SQPR}{S_{s s'}^{Q,PR}}

\newcommand{\SPRPR}{S_{s s'}^{PR,PR}}
\newcommand{\SPRQ}{S_{s s'}^{PR,Q}}

\newcommand{\gone}{\frac{E_{\mu}+E_{\nnu}}{(E_{\mu}+E_{\nnu})^{2}
                  -\omega^{2}(q)}}
\newcommand{\gtwo}{\frac{1}{(E_{\mu}+E_{\nnu})^{2}-\omega^{2}(q)}}

\def\rbra#1{(#1|}
\def\rket#1{|#1)}

\title{
Collective Paths Connecting the Oblate and Prolate Shapes
in $^{68}$Se and $^{72}$Kr 
Suggested by the Adiabatic Self-Consistent Collective Coordinate Method
}

\author{Masato KOBAYASI,$^1$~~ Takashi NAKATSUKASA,$^2$~~ 
Masayuki MATSUO$^3$ \\ 
and~ Kenichi MATSUYANAGI$^1$}

\inst{
{\small\it $^1$ Department of Physics, Graduate School of Science,}\\
{\small\it Kyoto University, Kitashirakawa, Kyoto 606-8502, Japan}\\
{\small\it $^2$ Institute of Physics and Center for Computational Science, 
University of Tsukuba,}\\
{\small\it Tsukuba 305-8571, Japan}\\
{\small\it $^3$ Graduate School of Science and Technology,}\\
{\small\it Niigata University, Niigata 950-2181, Japan}
}

\abst{
By means of the adiabatic self-consistent collective coordinate method
and the pairing-plus-quadrupole interaction, we have obtained
the self-consistent collective path connecting the oblate and prolate 
local minima in $^{68}$Se and $^{72}$Kr for the first time. 
The self-consistent collective path is found 
to run approximately along the valley
connecting the oblate and prolate local minima 
in the collective potential energy landscape. 
This result of calculation clearly indicates the importance of 
triaxial deformation dynamics in oblate-prolate shape coexistence phenomena.
}
\begin{document}

\maketitle

\section{Introduction}

Microscopic description of large amplitude collective motion in nuclei
is a long-standing fundamental subject of nuclear structure physics
\cite{rin80,bla86,kle91,dan00,kur01}.
In spite of the steady development in various theoretical concepts 
and mathematical formulations of them, application of the microscopic 
many-body theory to actual nuclear phenomena still remains as 
a challenging subject.$^{6)-33)}$
Shape coexistence phenomena are typical examples of large 
amplitude collective motion in nuclei, and
active investigations are currently going on 
both experimentally and theoretically.$^{34)-57)}$
We are particularly interested in the recent discovery of
coexisting two rotational bands in $^{68}$Se and $^{72}$Kr, which are
associated with oblate and prolate intrinsic shapes\cite{fis00,bou03}.
Clearly, these data strongly call for further development of the theory
which is able to describe them 
and renew our concepts of nuclear structure.
From the viewpoint of the microscopic mean-field theory,
coexistence of different shapes implies 
that different solutions of the Hartree-Fock-Bogoliubov (HFB) equations
(local minima in the deformation energy surface) appear 
in the same energy region and that the nucleus exhibits 
large amplitude collective motion connecting these different
equilibrium points. The identities and mixings of these different shapes
are determined by the dynamics of such collective motion.

On the basis of the time-dependent Hartree-Fock (TDHF) theory,
the selfconsistent collective coordinate (SCC) method was proposed
as a microscopic theory of such large amplitude collective motions
\cite{mar80}.
It was extended to the case of time-dependent HFB (TDHFB) 
including the pairing correlations\cite{mat86}, and
has been successfully applied to various kinds of anharmonic vibration 
and high-spin rotations
\cite{mat84,mat85a,mat85b,tak89,yam89a,yam89b,aib90,yam91,
ter91,ter92,mat92,shi01}.
In order to apply this method to shape coexistence phenomena, however,
we need to further develop the theory, since the known method of 
solving the basic equations of the SCC method,
called $\eta$-expansion method\cite{mar80}, assumes a single local minima
whereas several local minima of the potential energy surface 
compete in these phenomena. 
Some years ago, we proposed a new method of describing 
such large-amplitude collective motion, 
called the adiabatic self-consistent 
collective coordinate (ASCC) method\cite{mat00}.
This method provides a practical scheme of solving the basic equations 
of the SCC method\cite{mar80} 
using an expansion in terms of the collective momentum.
It does not assume a single local minimum, and therefore it is believed 
to be suitable for the description of shape coexistence phenomena. 
The ASCC method inherits the major achievement of the
adiabatic TDHF (ATDHF) methods and, in addition, 
enables us to include pairing correlations 
self-consistently. In this method,
the spurious number fluctuation modes are automatically
decoupled from the physical modes within a selfconsistent framework 
of the TDHFB theory.
This will be a great advantage when the method is applied to 
realistic nuclear problems.
To examine the feasibility of the ASCC method,
in Ref.~\citen{kob03}, we applied it to an exactly solvable model 
called the multi-$O(4)$ model\cite{mat82,miz81,suz88,fuk91}, 
which is a simplified version of 
the pairing-plus-quadrupole (P+Q) interaction model\cite{bar65,bar68,bes69}.
It was shown that this method yields a faithful 
description of tunneling motion through a barrier between
prolate and oblate local minima in the collective potential\cite{kob03}.

In this paper, we report on our first application of
the ASCC method to the P+Q interaction model. 
The major task here is to develop a practical procedure for
solving the basic equations of the ASCC method in order to
obtain the selfconsistent collective path.
We investigate, as typical examples, the oblate-prolate shape coexistence 
phenomena in $^{68}$Se and $^{72}$Kr\cite{fis00,bou03},
and find that the self-consistent collective paths run approximately 
along the valley connecting the oblate and prolate local minima 
in the collective potential energy landscape. 
To the best of our knowledge, this is the first time that,
starting from the microscopic P+Q Hamiltonian, 
the collective paths have been fully self-consistently obtained
for realistic situations,
although a similar approach to the study of 
large amplitude collective motion
was recently employed by Almehed and Walet\cite{alm04a,alm04b}. 

This paper is organized as follows: 
In \S2, the basic equations of the ASCC method are recapitulated. 
In \S3, we present a concrete formulation of
the ASCC method for the case of the P+Q Hamiltonian.
In \S4, an algorithm to solve the basic equations
of the ASCC method is discussed.
In \S5, we present results of numerical calculation for
the oblate-prolate shape coexistence phenomena in $^{68}$Se and $^{72}$Kr.
Concluding remarks are given in \S6.

A preliminary version of this work was reported in this journal\cite{kob04}.

\section{Basic equations of the ASCC method}

In this section, we summarize the basic equations of 
the ASCC method\cite{mat00}.
The basic assumption of our approach is that large-amplitude 
collective motions can be described by a set of time-dependent HFB state 
vectors $\ket\phiqppn$
parameterized by a single collective coordinate $q$,
the collective momentum $p$ conjugate to $q$,
the particle number $N$ and the gauge angle $\varphi$ conjugate to $N$.
Then, the state vectors can be written in the following form:
\begin{equation} 
\ket{\phiqppn}=e^{-i \varphi \Nhat}\ket{\phiqpn}
=e^{-i \varphi \Nhat}e^{ip{\hat Q(q)}}|\phi(q)\rangle.
\end{equation}
Making an expansion with respect to $p$ and requiring that 
the time-dependent variational principle be fulfilled 
up to the second order in $p$,
we obtain the following set of equations to
determine $\ket{\phiq}$, the infinitesimal generator
$\Qhat(q)$, and its canonical conjugate  $\Phat(q)$:
\vspace{\baselineskip}

\noindent
\underline{HFB equation in the moving frame}:
\begin{equation} 
\delta\bra{\phiq}\HhatMq \ket{\phiq} = 0,   
\label{eqcsmf}
\end{equation}
where
\begin{equation}
\HhatMq=\hat{H}-\lambda(q)\hat{N}-\frac{\partial V}{\partial q}\hat{Q}(q)
\end{equation}
represents the Hamiltonian in the moving frame. 
\vspace{\baselineskip}

\noindent
\underline{Local Harmonic equations in the moving frame}:
\begin{equation} 
\delta\bra{\phiq}[\HhatMq, \Qhat(q) ] - {1\over i} B(q) \Phat(q)
\ket{\phiq} = 0,   
\label{eqcshq}
\end{equation}
\begin{equation}
\delta\bra{\phiq} [\HhatMq, {1\over i}\Phat(q)] -C(q)\Qhat(q) 
-{1 \over 2B(q)}[[\HhatMq, (\Hhat - \lambda(q)\Nhat)_{A}], \Qhat(q)]
-{\del \lambda \over \del q}\Nhat
\ket{\phiq} = 0,  
\label{eqcshp}
\end{equation}
where
\begin{equation}
B(q) = -\bra{\phiq}[[\Hhat,\Qhat(q)],\Qhat(q)]\ket{\phiq}.
\end{equation}
represents the inverse mass,
\begin{equation}
C(q) = {\del^2 V \over \del q^2} 
+ {1\over 2B(q)}{\del B\over \del q}{\del V \over \del q}
\end{equation}
the local stiffness, and
$(\Hhat - \lambda\Nhat)_{A}$ 
denotes the two-quasiparticle creation and
annihilation parts of ($\Hhat - \lambda\Nhat$).

The infinitesimal generators, $\Qhat(q)$ and $\Phat(q)$, satisfy
\vspace{\baselineskip}

\noindent
\underline{Canonical variable condition}:
\begin{equation}
\bra{\phiq}[\Qhat(q),\Phat(q)]\ket{\phiq} = i. 
\label{cvcqp} 
\end{equation}
Once $\ket{\phiq}$ and the infinitesimal generators are determined
for every values of $q$,   
we obtain the collective Hamiltonian 
$\Hc(q,p) = {1\over 2} B(q) p^2 + V(q)$
with the collective potential
$V(q) = \bra{\phiq}\Hhat\ket{\phiq}$.  

In the above equations, no distinction is made between protons and
neutrons for simplicity of notation.
In the actual calculations described below, however, we shall explicitly
treat both the neutron number $N$ and the proton number $Z$.

\section{Application of the ASCC method to the P+Q model}

\subsection{The P+Q Hamiltonian and signature quantum number}

Let us start with the well-known P+Q Hamiltonian\cite{bar65,bar68,bes69},

\begin{equation}
{\hat H} = \HS - \sum_{\tau}\HPi - \HQQi.
\label{eqcshp0}
\end{equation}
\begin{eqnarray} 
A_{\tau}^{\dag} &=& {\sum_{k \in \tau}}'c_k^{\dag}c_{\tilde{k}}^{\dag},
~~~A_{\tau} = {\sum_{k \in \tau}}'c_{\tilde{k}}c_k,
\nonumber \\
D_{2K}   &=& \sum_{\tau=n,p} \sum_{kl \in \tau}  D_{2K}^{(\tau)}(kl) 
c_k^\dag c_l
\end{eqnarray}
Here,
$D_{2K}^{(\tau)}(kl) = \alpha_{\tau}^{2} \bra{k}r^2 Y_{2K}\ket{l}$;~
$G_{\tau}$ and $\chi$ denote the pairing and quadrupole force strengths, 
respectively; 
$c_k^{\dag}$ and $c_k$ are nucleon creation and annihilation operators
in the single-particle state $k$, 
while  $c_{\tilde{k}}^{\dag}$ and $c_{\tilde{k}}$
denote those in the time reversed state of $k$.
The index $\tau$ indicates protons ($\tau=p$) and neutrons($\tau=n$).
Although we shall not explicitly mention below, it should be keep in mind that
the single-particle index $k$ actually includes the index $\tau$.
The notation ${\rm \Sigma}'$ in the pair operators, $A_{\tau}^{\dag}$ and
$A_{\tau}$, indicates the sum over the pairs $(k, {\tilde{k}})$.  
The factors, $\alpha_n=(2Z/A)^{2/3}$ and $\alpha_p=(2N/A)^{2/3}$,
attached to the quadrupole matrix elements guarantee equivalent 
root-mean-square radii for protons and neutrons.
Following Baranger and Kumar\cite{bar68}, 
we take into account two major shells as a model space, and multiply
a reduction factor $\zeta=(N_L+3/2)/(N_L+5/2)$ to the quadrupole matrix
elements $D_{2K}^{(\tau)}(kl)$ of the upper harmonic-oscillator shell,
$N_L$ being the total oscillator quanta of the lower shell.
According to the conventional prescription 
of the P+Q interaction\cite{bar65,bar68,bes69},
we ignore the exchange (Fock) terms. Namely, we employ the
Hartree-Bogoliubov (HB) approximation 
throughout this paper.

We introduce the notations,
\begin{eqnarray}
{\hat F}_{s}^{(\pm)} &=&
\frac{1}{2}({\hat F}_{s}\pm{\hat F}_{s}^{\dag}),
\nonumber \\
{\hat F}_{s}^{(\pm)} &=& 
\{
A_{n}^{(\pm)}, A_{p}^{(\pm)}, D_{20}^{(\pm)}, D_{21}^{(\pm)}, D_{22}^{(\pm)} 
\}~~~~~(s=1-5),   
\end{eqnarray}
and write the P+Q Hamiltonian in the following form:
\begin{eqnarray} 
{\hat H} &=& \HS
  - \sum_{s=1}^5\frac{\kappa_{s}}{2} {\hat F}_{s}^{(+)}{\hat F}_{s}^{(+)}
  + \sum_{s=1}^5\frac{\kappa_{s}}{2} {\hat F}_{s}^{(-)}{\hat F}_{s}^{(-)},
\label{hpq}
\end{eqnarray}
where
$\kappa_s=\{ 2G_n, 2G_p, \chi, 2\chi, 2\chi \}$ for $s=1-5$.
Our Hamiltonian is invariant with respect to the rotation $\pi$ 
about the $x$ axis. The symmetry quantum number associated with it
is called signature, $r=e^{-i\pi\alpha}$. 
To exploit the signature symmetry,
it is convenient to use nucleon operators with definite signatures
defined by
\begin{eqnarray}
d_{k} &\equiv& \frac{1}{\sqrt{2}}(c_{k} + c_{\tilde{k}}),
~~~~~r=-i~(\alpha=1/2),
\nonumber \\
d_{{\bar k}} &\equiv& \frac{1}{\sqrt{2}}(c_{\tilde{k}} - c_{k}),
~~~~~r=+i~(\alpha=-1/2),
\label{sigbasis}
\end{eqnarray}
and their Hermite conjugates,  $d_{k}^{\dag}$ and  $d_{{\bar k}}^{\dag}$.
The operators ${\hat F}_{s}^{(\pm)}$ are then classified
according to the signature quantum numbers, $r=\pm 1$ ($\alpha=0,1$):
\begin{eqnarray}
  \{A_{n}^{(\pm)},A_{p}^{(\pm)},D_{20}^{(+)},D_{21}^{(-)},D_{22}^{(+)}\}
  ~~~~~ (r=+1),
  \nonumber \\
  \{D_{21}^{(+)},D_{22}^{(-)}\} ~~~~~    (r=-1).
\end{eqnarray}  
Note that $D_{20}^{(-)}= 0$.
The HB local minima corresponding to the oblate and prolate equilibrium shapes 
possess positive signature, $r=+1 (\alpha=0)$. Therefore, the 
operators, ${\hat Q}(q)$ and ${\hat P}(q)$, generating large amplitude
collective motions associated with these shapes also possess
positive signature. In other words, the negative signature degrees of freedom
are exactly decoupled from the large amplitude collective motion of interest,
so that we can ignore them.
Also, it is readily confirmed that the $K=1$ components associated with 
the quadrupole operator ${\hat D}_{21}^{(-)}$ exactly decouple from the
$K=0$ and 2 components in the local harmonic equations, (2.2) and (2.4).
As is well known, they are associated with the collective rotational motions,
and the large amplitude shape vibrational motions under consideration
are exactly decoupled from them in the present framework. 
We note, however, that it is possible, by a rather straightforward extension, 
to formulate the ASCC method in a rotating frame of reference.
By means of such an extension, we shall be able to take into account 
the coupling effects between the two kinds of large amplitude 
collective motion.  It is certainly a very interesting subject to study 
how the properties of the large-amplitude shape vibrational motions
change as a function of angular momentum, but it goes beyond the scope of 
this paper. We note, however, that some attempts 
toward this subject was recently made by Almehed and Walet\cite{alm04b}.

Thus, only the components, 
$\{A_{n}^{(\pm)},A_{p}^{(\pm)},D_{20}^{(+)},D_{22}^{(+)}\}$,
are pertinent to the shape coexistence dynamics of our interest.
They all belong to the positive signature sector, and 
we can adopt a phase convention with which single-particle
matrix elements of them are real. In the following, we assume that
this is the case.

\subsection{Quasiparticle-random-phase approximation (QRPA)
 at the HB local minima}

As discussed in the introduction, shape coexistence phenomena imply
the existence of several local minima in the deformation energy surface,
which are solutions of the HB equations. 
Let us choose one of them and denote it as  ${\ket{\phi_0}}$.
The HB equation is written
\begin{equation}
\delta{\bra{\phi_0}}{\hat H}
-\sum_{\tau}\lambda_{\tau} {\hat N}_{\tau}{\ket{\phi_0}}=0,
\label{HFB}
\end{equation}
where $\lambda_{\tau}$ denote chemical potentials for protons($\tau=p$)
and neutrons($\tau=n$). 
The quasiparticle creation and annihilation operators, $a_{\mu}^{\dag}$
and $a_{\mu}$, associated with the HB local minimum are 
defined by $a_{\mu}|\phi_0 \rangle=0$. Similar equations hold for
their signature partners $\nmu$. 
They are introduced through the Bogoliubov transformations:
\begin{equation}
\left(
 \begin{array}{c}
   a_{\mu}^{\dag} \\
   a_{\nmu}
 \end{array}
\right)
=
\sum_k
\left(
    \begin{array}{cc}
        U_{\mu k}&V_{\mu \nk}\\
        V_{\nmu k}&U_{\nmu \nk} 
    \end{array}
\right)
\left(
\begin{array}{c}
   d_{k}^{\dag}\\
   d_{\nk}
 \end{array}
\right),
\end{equation}
and their Hermite conjugate equations. 
(Here and hereafter, 
we do not mix protons and neutrons in these transformations.)
In terms of two quasiparticle creation and annihilation operators,
\begin{equation}
{\vect A_{\mu \nnu}^{\dag}} \equiv a_{\mu}^{\dag}a_{\nnu}^{\dag},~~~
{\vect A_{\mu \nnu}} \equiv a_{\nnu}a_{\mu},
\end{equation}
the RPA normal coordinates and momenta describing small amplitude
vibrations about the HB local minimum ${\ket{\phi_0}}$ are written
\begin{eqnarray}
{\Qhat}_{\rho} &=& 
\sum_{\mu \nnu} Q_{\mu \nnu}^{\rho}({\vect A_{\mu \nnu}^{\dag}}
+{\vect A_{\mu \nnu}}), \\
{\Phat}_{\rho} &=& 
i \sum_{\mu \nnu} P_{\mu \nnu}^{\rho}({\vect A_{\mu \nnu}^{\dag}}
-{\vect A_{\mu \nnu}}), 
\end{eqnarray}
where the sum is taken over the proton and neutron quasiparticle pairs
$(\mu \nnu)$,
and $\rho$ labels the QRPA modes.
The amplitudes, $Q_{\mu \nu}^{\rho}$ and $P_{\mu \nu}^{\rho}$,
are determined by the QRPA equations of motion,
\begin{equation} 
\delta\bra{\phi_0}[\Hhat-\sum_{\tau}\lambda_{\tau}\Nhat_{\tau},~\Qhat_{\rho} ] 
- {1\over i} B_{\rho} \Phat_{\rho}\ket{\phi_0} = 0,   
\end{equation}
\begin{equation}
\delta\bra{\phi_0} 
[\Hhat-\sum_{\tau}\lambda_{\tau}\Nhat_{\tau},~{1\over i}\Phat_{\rho}] 
-C_{\rho}\Qhat_{\rho} \ket{\phi_0} = 0,  
\end{equation}
and the orthonormalization condition,
$\bra{\phi_0}[\Qhat_{\rho},\Phat_{\rho'}]\ket{\phi_0} = i\delta_{\rho,\rho'}$.

\subsection{The HB equation and the quasiparticles in the moving frame}

For the P+Q Hamiltonian, the HB equation (\ref{eqcsmf})  
determining the state vector $|\phi(q)\rangle$ 
away from the local minimum reduces to
\begin{equation}
\delta\bra{\phiq}\hhatMq \ket{\phiq} = 0,
\label{mfHB}
\end{equation}
where $\hhatMq$ is the mean-field Hamiltonian in the moving frame,
\begin{eqnarray}
 \hhatMq &=& \hhatq  - \sum_{\tau}\lambda_{\tau}(q)\Nhat_{\tau}
                     - {\del V\over \del q}\Qhat(q), \\
   \hhatq &=& \sum_k \epsilon_k (d_k^{\dag}d_k + d_{\bar k}^{\dag}d_{\bar k}) 
            - \sum_s \kappa_s \Fhatp_s \bra{\phiq}\Fhatp_s\ket{\phiq}.
\end{eqnarray}
The state vector $|\phi(q)\rangle$
can be written as a unitary transform of $|\phi_{0} \rangle$:
\begin{eqnarray}
\ket{\phi(q)} &=& e^{{\hat \theta}(q)} \ket{\phi_0}, \nonumber \\
{\hat \theta}(q) &\equiv& 
\sum_{\mu \nnu}\theta_{\mu\nnu}(q)\Big(\amndag-\amn \Big),
\end{eqnarray}
where the sum is taken over the proton and neutron quasiparticle pairs
$(\mu \nnu)$.
The quasiparticle creation and annihilation operators,
$a_{\mu}^{\dag}(q)$ and $a_{\mu}(q)$, associated with
the state $|\phi(q)\rangle$,
which satisfy the condition, $a_{\mu}(q)|\phi(q) \rangle=0$,
are written: 
\begin{eqnarray}
a_{\mu}^{\dag}(q) &\equiv& 
e^{\hat \theta}(q) a_{\mu}^{\dag} e^{-{\hat \theta}(q)} 
 =\sum_{\nu} \Big( U_{\mu \nu}(q)a_{\nu}^{\dag} 
 +  V_{\mu \nnu}(q)a_{\nnu}\Big), \nonumber \\
a_{\nmu}(q) &\equiv& e^{\hat \theta}(q) a_{\nmu} 
e^{-{\hat \theta}(q)} 
 =\sum_{\nu} \Big(V_{\nmu \nu}(q)a_{\nu}^{\dag}+U_{\nmu \nnu}(q)a_{\nnu}\Big), 
\end{eqnarray}
where
\begin{equation}
\left(
    \begin{array}{cc}
        U_{\mu \nu}(q)&V_{\mu \nnu}(q)\\
        V_{\nmu \nu}(q)&U_{\nmu \nnu}(q) 
    \end{array}
\right)
= \left(
    \begin{array}{cc}
    \cos(\sqrt{\theta\theta^{T}})   & \displaystyle 
    -\theta\frac{\sin(\sqrt{\theta^{T}\theta})}{\sqrt{\theta^{T}\theta}}\\
    \displaystyle  \theta^{T}\frac{\sin(\theta \theta^{T})}
    {\sqrt{\theta \theta^{T}}}  & \cos({\sqrt{\theta^{T}\theta}}).
    \end{array}
\right) 
\end{equation}
Here, $\theta$ in the r.h.s. denotes the matrix 
composed of $\theta_{\mu\nu}(q)$,
and it is understood that its elements corresponding to those 
in the l.h.s. should be taken.

In terms of the quasiparticle operators defined above,
the mean-field Hamiltonian in the moving frame $\hat{h}_{M}(q)$,
the neutron and proton number operators, ${\hat N}_{\tau}$,
and the pairing and quadrupole operators, $\hat{F}_{s}^{(\pm)}$, 
are written in the following form:
\begin{eqnarray}
\hat{h}_{M}(q) &=& 
\bra\phiq \hat{h}_{M}(q) \ket\phiq
+\sum_{\mu}E_{\mu}(q)
\Big( {\vect{B}}_{\mu \mu}(q) + {\vect{B}}_{\nmu \nmu}(q) \Big), \\
\label{Fs}
\hat{N}_{\tau} &=& 
\bra\phiq \hat{N}_{\tau} \ket\phiq
+\sum_{\mu}N_{\tau}(\mu)\Big(\ammdag(q) + \amm(q)\Big)
\nonumber \\
&+& \sum_{\mu}N_{B,\tau}(\mu)
\Big( {\vect{B}}_{\mu \mu}(q) + {\vect{B}}_{\nmu \nmu}(q) \Big), \\
\hat{F}_{s}^{(\pm)} &=&
\bra\phiq \hat{F}_{s}^{(\pm)}\ket\phiq
+\sum_{\mu \nnu}F_{s}^{(\pm)}(\mu \nnu)\Big(\amndag(q) \pm \amn(q)\Big)
\nonumber \\
&+& \sum_{\mu \nu} F_{B,s}^{(\pm)}(\mu \nu) 
\Big({\vect{B}}_{\mu \nu}(q) + {\vect{B}}_{\nmu \nnu}(q)\Big), 
\label{Fhatp}
\end{eqnarray}
where
\begin{equation}
{\vect A_{\mu \nnu}^{\dag}}(q) \equiv a_{\mu}^{\dag}(q)a_{\nnu}^{\dag}(q),~~
{\vect A_{\mu \nnu}}(q) \equiv a_{\nnu}(q)a_{\mu}(q),~~
{\vect B_{\mu\nu}}(q) \equiv a_{\mu}^{\dag}(q)a_{\nu}(q).
\end{equation}
Note that $E_{\nmu}(q)=E_{\mu}(q)$ 
and also that the equalities, 
$F_{B,s}^{(\pm)}(\nmu \nnu) = F_{B,s}^{(\pm)}(\mu \nu)$,
hold for the operators under consideration.                    
Explicit expressions for the expectation values
and the quasiparticle matrix elements
appearing in the above equations are given in Appendix A.

\subsection{Local harmonic equations in the moving frame}

We can represent the infinitesimal generators, $\hat{Q}(q)$ and $\hat{P}(q)$, 
in terms of $\amndag(q)$ and $\amn(q)$ as
\begin{eqnarray}
{\hat Q(q)} &=& \sum_{\mu \nnu} Q_{\mu \nnu}(q)\Big(\amndag(q)+\amn(q)\Big), \\
{\hat P(q)} &=& i \sum_{\mu \nnu} P_{\mu \nnu}(q)\Big(\amndag(q)-\amn(q)\Big),
\end{eqnarray}
where the sum is taken over the proton and neutron quasiparticle pairs 
$(\mu \nnu)$.
For the P+Q Hamiltonian, 
the local harmonic equations, (\ref{eqcshq}) and (\ref{eqcshp}), 
in the moving frame reduce to
\begin{equation} \label{sephq}
\delta\bra{\phiq}[\hhatMq, \Qhat(q) ] 
- \sum_s f^{(-)}_{Q,s} \Fhatm_s 
- {1\over i} B(q) \Phat(q) 
     \ket{\phiq} = 0,
\end{equation}
\begin{eqnarray} \label{sephp}
\delta\bra{\phiq}[\hhatMq&,& {1\over i}B(q)\Phat(q)] 
    - \sum_s f^{(+)}_{P,s} \Fhatp_s 
    - B(q)C(q)\Qhat(q) 
    - \sum_s f^{(+)}_{R,s} \Fhatp_s \nonumber \\ 
    &+& \sum_s f^{(-)}_{Q,s} \Rhat_s^{(-)} 
    -\sum_{\tau}f_{N,{\tau}}\Nhat_{\tau}
    \ket{\phiq} =0, 
\end{eqnarray}
where the quantities $f^{(-)}_{Q,s}$ etc. are given by
\begin{eqnarray}
 f^{(-)}_{Q,s} &\equiv& - \kappa_s 
              \bra{\phiq}[\Fhatm_s, \Qhat(q)] \ket{\phiq}
              =2{\ks}( {\FAsmni},Q(q) ), \\
 f^{(+)}_{P,s} &\equiv& 
   \kappa_s \bra{\phiq}[\Fhatp_s,{1\over i}B(q)\Phat(q)] \ket{\phiq}
   =2{\ks}B(q)( {\FAspni},P(q) ), \\
 f^{(+)}_{R,s} &\equiv& -\frac{1}{2}\kappa_s 
 \bra{\phiq}[\Rhat_s^{(+)},\Qhat(q)] \ket{\phiq}
   = {\ks}({\RAspni},Q(q) ) , \\
 f_{N,{\tau}} &\equiv& B(q){\del \lambda_{\tau} \over\del q}.
\end{eqnarray}
Here we have introduced the notations, 
\begin{equation}
 \Rhat_s^{(\pm)}  \equiv [ \Fhat_{B,s}^{(\pm)}, (\hhat(q)
 -\sum_{\tau}\lambda_{\tau}(q)\Nhat_{\tau})_{A}] 
 \equiv \sum_{\mu \nnu }R_s^{(\pm)}(\mu \nnu) \Big(\amndag(q) \mp \amn(q)\Big),
\end{equation}
where $(\hhat(q)-\sum_{\tau}\lambda_{\tau}(q)\Nhat_{\tau})_{A}$ represents
the ${\vect A_{\mu \nnu}^{\dag}}(q)$ and ${\vect A_{\mu \nnu}}(q)$ parts
of the operator in the parenthesis.
We also use the notations,
\begin{equation}
(F_s^{(-)}, Q(q)) \equiv \sum_{\mu\nnu} F_s^{(-)}(\mu\nnu)Q_{\mu\nnu}(q),
~~~{\rm etc}.
\end{equation}
Note that $f^{(-)}_{Q,s}, f^{(+)}_{P,s}$ and $f^{(+)}_{R,s}$ are linear
functions of $Q_{\mu\nnu}(q)$ or $P_{\mu\nnu}(q)$.

One can easily derive the following expressions for the matrix elements,
$Q_{\mu \nnu}(q)$  and $P_{\mu \nnu}(q)$,
from the local harmonic equations in the moving frame, 
(\ref{sephq}) and (\ref{sephp}): 
\begin{eqnarray}
Q_{\mu \nnu}(q) &=& \sum_{s} g_1(\mu\nnu){\FAsmni}(\mu \nnu){\fq}
 + \sum_{s}g_2(\mu\nnu)\big\{ {\FAspni}(\mu \nnu){\fpr}   \nonumber \\
&+&{\RAsmni}(\mu \nnu){\fq}+\sum_{\tau}N_{\tau}(\mu\nnu)f_{N,\tau} \big\}   \\
\label{qi}
P_{\mu \nnu}(q) &=& \sum_{s}g_1(\mu\nnu)\big\{ {\FAspni}(\mu \nnu){\fpr} 
 + {\RAsmni}(\mu \nnu){\fq}+\sum_{\tau}N_{\tau}(\mu\nnu)f_{N,\tau} \big\}    
 \nonumber \\
&+&{\omega^{2}(q)}\sum_{s}g_2(\mu\nnu){\FAsmni}(\mu \nnu){\fq}, 
\label{pi}
\end{eqnarray}
where $f_{PR,s}^{(+)}=f_{P,s}^{(+)}+f_{R,s}^{(+)}$
and 
\begin{equation}
g_{1}(\mu \nnu) \equiv \gone, ~~~
g_{2}(\mu \nnu) \equiv \gtwo. 
\end{equation}
Note that $\omega^2$, representing the square of the 
frequency $\omega(q)=\sqrt{B(q)C(q)}$ of the local harmonic
mode, is not necessarily positive.
The values of $B(q)$ and $C(q)$ depend on the scale of
the collective coordinate $q$, while $\omega(q)$ does not.
Namely, the scale of $q$ can be chosen arbitrarily 
without affecting the frequency $\omega(q)$.
We thus require $B(q)=1$ everywhere on the collective path
to uniquely determine the scale of $q$.

Inserting expressions (3.34) and (3.35) for $Q_{\mu \nnu}(q)$ and
$P_{\mu \nnu}(q)$ into Eqs.~(3.28)-(3.30), and combining them with
the orthogonality condition to the number operators, 
\begin{equation}
\langle\phi(q)|[\hat{N}_{\tau},\hat{P}(q)]|\phi(q)\rangle 
          = 2i( P(q),N_{\tau} ) = 0,
\end{equation}
we obtain linear homogeneous equations:
\begin{equation}
\sum_{s'\tau '}
\left(
    \begin{array}{ccc}
       S_{ss'}^{Q,Q}     &  S_{ss'}^{Q,PR}     &  S_{s\tau '}^{Q,N}      \\
                        &                     &                        \\
       S_{ss'}^{PR,Q}   &  S_{ss'}^{PR,PR}    &  S_{s\tau '}^{PR,N}    \\
                        &                     &                        \\  
       S_{\tau s'}^{N,Q} &  S_{\tau s'}^{N,PR} &  S_{\tau \tau '}^{N,N} 
    \end{array}
\right)
\left(
\begin{array}{c}
   f_{Q,s'}^{(-)}  \\
                           \\
   f_{PR,s'}^{(+)} \\
                           \\
   f_{N,\tau '}
 \end{array}
\right)
=0,
\label{dispeq0}
\end{equation}
for the vectors,
${\vect f}_{Q}^{(-)}$,~${\vect f}_{PR}^{(+)}$, and ${\vect f}_{N}$,
defined by
\begin{eqnarray}
{\vect f}_{Q}^{(-)} &\equiv& \{f_{Q,1}^{(-)},~f_{Q,1}^{(-)}\},\\
{\vect f}_{PR}^{(+)} &\equiv& 
          \{f_{PR,1}^{(+)},~f_{PR,2}^{(+)},~f_{PR,3}^{(+)},~f_{PR,5}^{(+)}\},\\
{\vect f}_{N} &\equiv& \{f_{N,n},~f_{N,p}\}.
\end{eqnarray}
Here,
\begin{eqnarray}
\SQQ &\equiv&   2({\FAsmni},{\FAsmpni})_{g_{1}}
        +2 ({\FAsmni},{\RAsmpni})_{g_{2}} - \frac{1}{\kappa_s}\delta_{ss'}, \\
\SQPR &\equiv&  2({\FAsmni},{\FAsppni})_{g_{2}}, \\
S_{s\tau '}^{Q,N} &\equiv& 2({\FAsmni}, N_{\tau '})_{g_{2}}, \\
\SPRQ &\equiv&  2({\FAspni},{\RAsmpni})_{g_{1}}
               +2\omega^{2}(q)({\FAspni},{\FAsmpni})_{g_{2}}
               +({\RAspni},{\RAsmpni})_{g_{2}}, \nonumber \\
              &+&({\RAspni},{\FAsmpni})_{g_{1}}, \\
\SPRPR &\equiv& 2({\FAspni},{\FAsppni})_{g_{1}}
         +({\RAspni},{\FAsppni})_{g_{2}}- \frac{1}{\kappa_s}\delta_{ss'}, \\
S_{s\tau '}^{PR,N} &\equiv&  2({\FAspni},N_{\tau '})_{g_{1}}
                +({\RAspni},N_{\tau '})_{g_{2}}, \\
S_{\tau s'}^{N,Q} &\equiv&   \omega^{2}(q)(N_{\tau}, F_{s'}^{(-)})_{g_{2}}
                +(N_{\tau}, R_{s'}^{(-)}   )_{g_{1}}, \\
S_{\tau s'}^{N,PR} &\equiv&  (N_{\tau}, F_{s'}^{(+)})_{g_{1}}, \\
S_{\tau\tau'}^{N,N} &\equiv&  (N_{\tau}, N_{\tau'} )_{g_{1}},
\end{eqnarray}
with the notations,
\begin{equation}
(F_s^{(-)},F_{s'}^{(-)})_{g_{1}} \equiv 
 \sum_{\mu\nnu}F_{s}^{(-)}(\mu\nnu)g_{1}(\mu\nnu) F_{s'}^{(-)}(\mu\nnu),
 ~~~{\rm etc}.
\end{equation}
Equation (\ref{dispeq0}) is of the form
\begin{equation}
\sum_{\sigma'=1}^8 S_{\sigma \sigma'}(\omega^{2}(q)) f_{\sigma'}=0,
\label{dispersion}
\end{equation}
for the vector ${\vect f}$ composed of
\begin{eqnarray}
 \{f_{\sigma=1-8}\} &\equiv& 
\{ {\vect f}_{Q}^{(-)},~{\vect f}_{PR}^{(+)},~{\vect f}_{N}\} \\
 &\equiv& \{ f_{Q,1}^{(-)},~f_{Q,2}^{(-)},~f_{PR,1}^{(+)},~
 f_{PR,2}^{(+)},~f_{PR,3}^{(+)},~f_{PR,5}^{(+)},~f_{N,n},~f_{N,p} \},
\end{eqnarray}
so that the frequency $\omega$ of the local harmonic mode is determined 
by the condition $\det S=0$.  
The normalizations of ${\vect f}$ are fixed by  
\begin{equation}
\langle\phi(q)|[\hat{Q}(q),\hat{P}(q)]|\phi(q)\rangle 
          = 2i( Q(q),P(q) ) = i.
\end{equation}
Note that
$\omega^2$ represents the curvature of the collective potential,
\begin{equation}
\omega^2=\frac{\partial^2 V}{\partial q^2}, 
\label{freq}
\end{equation}
for the choice of coordinate scale with which $B(q)=1$.

In concluding this section, we mention that the reduction of the local
harmonic equations to linear homogeneous equations like (\ref{dispersion})
can be done for any effective interaction that can be written as a sum of
separable terms.
Below, we call the local harmonic equations in the moving frame
``moving frame QRPA" for brevity.

\section{Procedure of calculation}

\subsection{Algorithm to find collective paths}

In order to find the collective path connecting the oblate and prolate
local minima, we have to determine the state vectors $\ket{\phiq}$
and the infinitesimal generators, $\Qhat(q)$ and $\Phat(q)$,
by solving the moving frame HB equation (\ref{mfHB}) and 
the moving frame QRPA equations, (\ref{sephq}) and (\ref{sephp}).
Since $\Qhat(q)$ depends on $\ket{\phiq}$ and vice versa,
we have to resort to some iterative procedure.
We carry out this task through the following algorithm.

Let us assume that the state vector, $\ket{\phiq}$, 
and the infinitesimal generators, $\Qhat(q)$ and $\Phat(q)$, 
are known at a specific point of $q$. 
We then find the state vector, $|\phi(q+\delta q)\rangle$, 
and the infinitesimal generators, 
$\Qhat(q+\delta q)$ and $\Phat(q+\delta q)$, 
in the neighboring point, $q+\delta q$, through the following steps.

\noindent
~{\it Step 1}:~~Construct a state vector at the neighboring point, $q+\delta q$,
with the use of $\Phat(q)$,
\begin{equation}
|\phi(q+\delta q)\rangle^{(0)} = e^{-i \delta q {\hat P(q)}}|\phi(q)\rangle.
\label{deltaq}
\end{equation}
Though $|\phi(q+\delta q)\rangle^{(0)}$ does not necessarily satisfy
the moving frame HB equation, (\ref{mfHB}), 
we can use this state vector as an initial guess at $q+\delta q$.\\ 
~{\it Step 2}:~~Solve the moving frame HB equation (\ref{mfHB}) 
with the use of $\Qhat^{(0)}(q+\delta q)=\Qhat(q)$ 
as an initial guess for $\Qhat(q+\delta q)$,
and obtain an improved state vector $|\phi(q+\delta q)\rangle^{(1)}$.
In doing this, we find it important to impose the constraint,
\begin{equation}
 \langle\phi(q+\delta q)|\Qhat(q)|\phi(q+\delta q)\rangle = \delta q
 \label{constr2}
\end{equation}
for the increment $\delta q$ of the collective coordinate $q$,
together with the constraints,
\begin{eqnarray}
 \langle\phi(q+\delta q)|\Nhat_{\tau}|\phi(q+\delta q)\rangle &=& N_{\tau},
 \quad\quad \tau=p,n
 \label{constr1} 
\end{eqnarray}
for the proton and neutron numbers $(N_p=Z,~N_n=N)$.
The constraint (\ref{constr2}) is easily derived by combining 
Eq.~(\ref{deltaq}) with the canonical variable condition (\ref{cvcqp}).
Details of this step is described in Appendix B. \\
~{\it Step 3}:~~Solve the moving frame QRPA equations,
(\ref{sephq}) and (\ref{sephp}),
with the use of $|\phi(q+\delta q)\rangle^{(1)}$ 
to obtain $\hat Q^{(1)}(q+\delta q)$ and  $\hat P^{(1)}(q+\delta q)$.\\ 
~{\it Step 4}:~~Go to {\it Step 2}
and solve Eq. (\ref{mfHB}) with the use of $\hat Q^{(1)}(q+\delta q)$.

If the iterative procedure, {\it Steps 2-4}, converges, 
we obtain the selfconsistent 
solutions, $\Qhat(q+\delta q)$, $\Phat(q+\delta q)$ and 
$|\phi(q+\delta q)\rangle$, that satisfy 
Eqs. (\ref{mfHB}), (\ref{sephq}) and (\ref{sephp}) 
simultaneously at $q+\delta q$.
Then, we go to {\it Step 1} to construct 
an initial guess $|\phi(q+2\delta q)\rangle^{(0)}$ 
for the next point, $q+2\delta q$, and repeat the above
procedure. In this way, we proceed step by step along the collective path.

The above is a brief summary of the basic algorithm. 
In actual numerical calculations,
we start the procedure from one of the HB local minima and 
choose the lowest frequency QRPA mode as an initial condition
for the infinitesimal generators, $\Qhat$ and $\Phat$, at $q=0$. 
Under ordinary conditions, 
we can proceed along the collective path following 
the procedure described above. In some special situations, however,
we need additional considerations concerning the choice of the
initial guess, $\Qhat^{(0)}(q+\delta q)$, in {\it Step 2}.
Actually, we meet such situations in some special regions of 
the collective path in $^{72}$Kr. We shall give a detailed discussion 
on this point in $\S$\S~5.3.

We have checked that the same collective path is
obtained by starting from the other local minimum and 
proceeding in an inverse way.

\subsection{Details of calculation}

In the numerical calculation, we use the spherical single-particle energies 
of the modified oscillator model of Ref.~\citen{ben85}, 
which are listed in Table \ref{table:1},
and follow the conventional prescriptions
of the P+Q interaction model\cite{bar68}, except that 
the pairing and quadrupole interaction strengths, $G_{\tau}$ and $\chi$,
are chosen to approximately reproduce the pairing gaps and 
quadrupole deformations obtained in the Skyrme-HFB calculation 
by Yamagami {\it et al.}\cite{yam01}.
Their values are: $G_n=0.320~(0.299)$ $G_p=0.320~(0.309)$ 
and $\chi' \equiv \chi b^4 = 0.248~(0.255)$ 
in units of MeV for $^{68}$Se~($^{72}$Kr),
where $b$ is the length parameter given by
$b^2=\frac{4}{5}\big(\frac{2}{3}\big)^{1/3}r_{0}^{2}A^{1/3}$. 
The pairing gaps, $\Delta_{\tau=p,n}$, and 	
deformation parameters, $\beta$ and $\gamma$, are defined as usual 
through the expectation values of the pairing and quadrupole operators:
\begin{eqnarray}
\Delta_{\tau}(q) &=& G_{\tau}\matel{\phi(q)}{\PAIR}{\phi(q)},\\
\beta \cos{\gamma} &=& 
\chi' \matel{\phi(q)}{{\hat D}_{20}^{(+)}}{\phi(q)}
                                                   /(\hbar \omega_0 b^2),\\
\beta \sin{\gamma} &=& 
\sqrt{2} \chi' \matel{\phi(q)}{{\hat D}_{22}^{(+)}}{\phi(q)}
                                                   /(\hbar \omega_0 b^2),
\end{eqnarray}
where $\hbar\omega_0$ denotes the frequency  
of the harmonic oscillator potential.

\begin{table}
\begin{center}
\caption{
Spherical single-particle orbits and their energies used in the calculation.
The energies relative to those of $1g_{9/2}$ are given in units of MeV.
}
\label{table:1}   
\begin{tabular}{cccccccccc} \hline \hline
orbits 
& $1f_{7/2}$ &$2p_{3/2}$ & $1f_{5/2}$   
& $2p_{1/2}$ &$1g_{9/2}$ & $2d_{5/2}$ 
& $1g_{7/2}$ &$3s_{1/2}$ & $2d_{3/2}$ \\ \hline
protons    
&  -8.77     &  -4.23    & -2.41  
&  -1.50     &   0.0     &  6.55  
&   5.90     &  10.10    &  9.83  \\
neutrons   
&  -9.02     &  -4.93    & -2.66  
&  -2.21     &   0.0     &  5.27  
&   6.36     &   8.34    &  8.80  \\ \hline
\end{tabular}
\end{center}
\end{table}

\section{Results of calculation}

\subsection{Properties of the QRPA modes at the local minima 
in $^{68}$Se and $^{72}$Kr}

For the P+Q Hamiltonian described in $\S$ 4, 
the lowest HB solution corresponds to the oblate shape, 
while the second lowest HB solution possesses 
the prolate shape, for both $^{68}$Se and $^{72}$Kr
(see Table~\ref{table:2}).
Their energy differences are 0.30 and 0.82 MeV for
$^{68}$Se and $^{72}$Kr, respectively.
In the QRPA calculations at these local minima,
we obtain strongly collective quadrupole modes with low-frequencies.
They correspond to the $\beta$ and $\gamma$
vibrations in deformed nuclei with axial symmetry. 
Although the former in fact contains pairing vibrational components,
we call it $\beta$ vibration because the transition matrix elements
for the quadrupole operator $D_{20}^{(+)}$ are enhanced.
(A neutron pairing vibrational mode appears as the second QRPA mode
at the oblate minimum in  $^{72}$Kr; see Table~\ref{tab:2}.)
Let us note that there is an important difference between 
$^{68}$Se and $^{72}$Kr
concerning the relative excitation energies of the $\beta$ and $\gamma$
vibrational modes:
In the case of $^{68}$Se, the frequencies of the $\gamma$ vibrational
QRPA mode are lower than those of the $\beta$ vibrational one
for both the oblate and prolate local minima.
The situation is opposite in the case of $^{72}$Kr,
namely, the $\beta$ vibrations are lower than the $\gamma$ vibrations.
As we shall see in the succeeding subsections, 
this difference leads to an important difference in the properties
of the collective path connecting the two local minima.

\begin{table}
\begin{center}
\caption{
The equilibrium quadrupole deformation parameters ($\beta$, $\gamma$),
the pairing gaps ($\Delta_{\tau}$) in units of MeV,
the QRPA eigenenergies $\hbar\omega_{\rho=1,2}$ in units of MeV,
and the relevant quadrupole transition matrix elements squared, 
$|M_{\rho}|^2 \equiv |\langle\rho|D_{2K}^{(+)}|0\rangle|^2$
($\rho=1,2$).
Here, $|\rho\rangle$ and $|0\rangle$ denote the QRPA one-phonon 
and the ground states.
The notations, $(\beta)$, $(\gamma)$, and $(\Delta_n)$, 
in the eighth column, respectively, indicate 
the $\beta$-, the $\gamma$- and the neutron-pairing vibrational modes;
the $|M_{\rho}|^2$ values for $K=0,~2$, and 0 
are presented, respectively, in the Weisskopf units. 
}    
\label{table:2}
\begin{tabular}{ccccccccc} \hline \hline
& $\beta$ & $\gamma$ & $\Delta_n$ 
& $\Delta_p$ & $\omega_1$ & $|M_1|^2$
& $\omega_2$ & $|M_2|^2$ 
\\ \hline
$^{68}$Se~(prolate)     
&  0.234     &   $0^{\circ}$      &  1.34  
&   1.42     &  1.02($\gamma$)    & 33.66   
&   1.91($\beta$)  & 12.19           \\
$^{68}$Se~(oblate)    
&  0.284     &   $60^{\circ}$     &  1.17  
&   1.27     &  1.55($\gamma$)    & 13.64    
&   2.25($\beta$)   &  7.67          \\
$^{72}$Kr~(prolate)    
&  0.376     &   $0^{\circ}$      &  1.15  
&   1.29     &  1.60($\beta$)     &  12.97   
&   1.67($\gamma$)   & 14.61           \\
$^{72}$Kr~(oblate)    
&  0.354     &   $60^{\circ}$     &  0.86  
&   1.00     &  1.15($\beta$)     &  5.37   
&   1.91($\Delta_n$)   &  0.19         \\
\hline
\end{tabular}
\end{center}
\end{table}

\subsection{Collective path connecting the oblate and prolate minima 
in $^{68}$Se}

As the $\gamma$ vibrational mode is the lowest-frequency and most collective
QRPA mode at the prolate local minimum, we have chosen this mode
as an initial condition for solving the basic equations of the ASCC method, 
and carried out the procedure described in \S\S~4.1. 
We thus obtained the collective path connecting the oblate and prolate 
local minima in \Se, which is plotted in Fig.~1(a).
As we have extracted the collective path in the TDHB phase space,
which has huge degrees of freedom, the path drawn in this figure should be 
regarded as a projection of the collective path onto the ($\beta, \gamma$)
plane.
Roughly speaking, the collective path goes through 
the valley that exists in the $\gamma$ direction
and connects the oblate and prolate minima. 
If $\beta$ is treated as a collective coordinate and
the oblate and prolate shapes are connected through the spherical point,
the variation of the potential energy would be much greater 
than that along the collective path we obtained.
The potential energy curve $V(q)$ along the collective path
evaluated using the ASCC method is shown in Fig.~1(b).
Since we have defined the scale of the collective coordinate $q$ 
such that the collective mass is given by $M(q)=B(q)^{-1}=1$~MeV$^{-1}$, 
the collective mass as a function of the geometrical length $s$ 
along the collective path in the $(\beta,\gamma)$ plane can be defined by
\begin{equation}
 M(s(q))=M(q)(\frac{ds}{dq})^{-2} 
\end{equation}
with $ds^2=d\beta^2+\beta^2d\gamma^2$.
This quantity is presented in Fig.~1(c) as a function of $q$.
The triaxial deformation parameter $\gamma$ is plotted 
as a function of $q$ in Fig.~1(d).
Variations of the pairing gaps, $\Delta_{\tau}(q)$, 
and of the eigen-frequencies of the moving frame QRPA equations 
along the collective path are plotted in Figs.~1(e) and 1(f). 
The solid curve in Fig.~1(f) represents the frequency squared,
$\omega^{2}(q)=B(q)C(q)$,
given by the product of the inverse mass $B(q)$ and
the local stiffness $C(q)$, of the solutions of 
the moving frame QRPA equations, 
that corresponds to the $\gamma$-vibration
in the oblate and prolate limits.
These QRPA solutions determine the infinitesimal generators
$\Qhat(q)$ and $\Phat(q)$ along the collective path.
For reference sake, we also present in Fig.~1(f) another
solution of the moving frame QRPA equations, which
possesses the $\beta$-vibrational character and is irrelevant to
the collective path in the case of $^{68}$Se.
Note that the frequency of the $\gamma$-vibrational mode becomes imaginary
in the region, $12^{\circ} < \gamma < 45^{\circ}$.
These results should reveal interesting dynamical properties of 
the shape coexistence phenomena in $^{68}$Se. 
For instance, the large collective mass
in the vicinity of $\gamma=60^\circ$ (Fig. 1(c)) might increase 
the stability of the oblate shape in the ground state. 
Detailed investigation of these quantities  
as well as solutions of the collective Schr\"odinger equation 
will be given in the succeeding paper\cite{kob05}.

\subsection{Collective path connecting the oblate and prolate minima 
in $^{72}$Kr}

In contrast to $^{68}$Se,
the lowest-frequency QRPA mode is the $\beta$ vibration
at the prolate local minimum in $^{72}$Kr.
Therefore, we have chosen this mode
as the initial condition at the prolate minimum
and started the procedure of extracting the collective path.  
Then, the collective path first goes in the direction of 
the $\beta$ axis on the $(\beta,\gamma)$ plane.
As we go along the $\beta$ axis, we eventually encounter 
a situation that the two solutions of the moving frame QRPA equations
compete in energy, and they eventually cross each other. 
Namely, the character of the solution 
with the lowest value of $\omega^2=BC$ changes from the
$\beta$ vibrational to the $\gamma$ vibrational ones
at some point on the collective path.
If only the solution $\Qhat_1(q)$ with the lowest value of $\omega^2$
at the previous point $q$ is always chosen as an initial guess
for $\Qhat(q+\delta q)$
in {\it Step 2} of the algorithm described in $\S\S$~4.1,
then the direction of the collective path on the $(\beta,\gamma)$ plane
change abruptly from the $\beta$ to the $\gamma$ directions
immediately after the crossing point 
(in the vicinity of the point C' in Fig.~3 presented below),  
and the numerical algorithm outlined in $\S\S$~4.1 fails
at this point:
During the iterative procedure of solving the moving frame HB equation, 
we encounter a situation where 
the overlap $(Q(q), Q(q+\delta q))$ between the infinitesimal
generators $\hat{Q}$ at the neighboring points, $q$ and $q+\delta q$,
vanishes, because the former possesses $K$=$0$ whereas the latter has $K$=$2$.
The numerical algorithm (details of which is described in Appendix B) 
then stops to work just at this point,
where the overlaps $(N_{\tau}, Q(q+\delta q))$ also vanish by the same reason.
This problem occurs even if we decrease the step size $\delta q$.
We find, however, that we can avoid this difficulty 
by employing a more suitable initial guess for $\Qhat(q+\delta q)$.
Namely, we take a linear combination of the two solutions,
$\Qhat_1(q)$ and $\Qhat_2(q)$ at the previous point $q$,
$\Qhat^{(0)}(q+\delta q)=(1-\varepsilon)\Qhat_1(q)+\varepsilon \Qhat_2(q)$,
with a small coefficient $\varepsilon$, as an initial guess.
This improvement is just for starting 
the iterative procedure at the next point, $q+\delta q$, 
on the collective path, 
so that the self-consistent solution, $\Qhat(q+\delta q)$,
obtained upon convergence of the iterative procedure, 
of course, do not depend on the values of $\varepsilon$.
For instance, we obtain an axially symmetric solution $\ket{\phi(q+\delta q)}$
and a generator $\Qhat(q+\delta q)$ preserving the $K$ quantum number
in the region with $\beta > 0.24$ around the prolate minimum,
even when we start the iterative procedure using 
an initial guess for $\Qhat(q+\delta q)$ that breaks the axial symmetry.
We confirmed that this is indeed the case 
as long as $\varepsilon$ is a small finite value around 0.1.
This special care is needed only near such crossing points
(as shown below in Figs.~3-5)
where two solutions of the moving frame QRPA equations 
with different $K$ quantum numbers compete in energy.

With the improved algorithm mentioned above,
we have successfully obtained the smooth deviation of the direction 
of the collective path from the $\beta$ axis
toward the $\gamma$ direction (see Fig.~2(a)). 
We note that the character of the lowest $\omega^2$ solution of 
the local harmonic equations also gradually changes from the 
$\beta$ vibrational to the $\gamma$ vibrational ones (see Fig.~2(f)). 
The details of the turn over region is presented in Fig.~3.
One can clearly see in Fig.~3(a) a gradual onset of axial-symmetry breaking
in the solutions $\ket{\phiq}$ of the moving frame HB equation.
One can also see in Fig.~3(b) an avoided crossing between 
the lowest two solutions of the moving frame QRPA equations 
associated with mixing of the components with $K$=$0$ and 2.
After the smooth turn to the $\gamma$ direction, the $\gamma$ value
increases keeping the $\beta$ value roughly constant, 
and the collective path eventually approaches to the $\gamma=60^{\circ}$ axis.
Then, we again encounter a similar situation.
Adopting the improved algorithm, 
we have confirmed that the character of the lowest $\omega^2$ solution 
smoothly changes, this time, from the $\gamma$ vibrational to 
the $\beta$ vibrational ones.
The collective path thus merges with the $\gamma=60^{\circ}$ axis, and
finally reaches the oblate minimum.  

We have also carried out the calculation 
starting from the oblate minimum and proceeded in an inverse way,
and obtained the same collective path.
This should be regarded as a crucial test of consistency of our calculation.
Figure 4 shows the details of this test:
The collective path that started from the prolate minimum and 
turned into the $\gamma$ direction
gradually merges with the $\gamma=60^{\circ}$ axis.
On the other hand, moving in the opposite direction,
we see a gradual onset of axial-symmetry breaking in the collective path
that started from the oblate minimum.
We see that the two results of calculation for the collective path 
nicely agree with each other.
The importance of taking account of the mixing between 
the $\beta$- and $\gamma$-vibrational degrees of freedom 
in solving the moving frame HB and QRPA equations 
is again demonstrated in Fig.~5, which shows the details of
the turn over region from the $\gamma=60^{\circ}$ axis.

Although the collective path drawn in Fig.~2(a) should be 
regarded as a projection of it onto the $(\beta,\gamma)$ plane,
the result of calculation indicates that the collective path runs 
roughly along the valley in this plane.
The potential energy curve $V(q)$, the collective mass $M(s(q))$,
variations of the pairing gaps, $\Delta_{\tau}(q)$,
are presented in Figs.~2(b),~(c),~(e), respectively. 
Their properties are similar to those for $^{68}$Se.
In particular, we notice again a significant increase of $M(s(q))$ 
in the vicinity of the oblate minimum.  

Quite recently, Almehed and Walet studied the oblate-prolate shape 
coexistence phenomenon in $^{72}$Kr by means of an approach 
similar to the ASCC method but with some additional approximations\cite{alm04b},
and found a collective path going from the oblate minimum over a spherical
energy maximum into the prolate secondary minimum.
We have also obtained such a collective path when we impose the
axially symmetry to the solutions $\ket{\phiq}$ of the moving frame HB equation 
and always use only $K$=$0$ solutions of the moving frame QRPA equations. 
But, when we relax such symmetry restrictions and follow the lowest $\omega^2$
solution of the moving frame QRPA equations, we obtain the
collective path presented in Fig.~2, which breaks the axial symmetry.  
The reason of this disagreement is not clear at present.
With their parameters of the P+Q Hamiltonian, 
they did not encounter the character change of the
lowest moving-frame-QRPA mode on the collective path,
from the $\beta$ vibrational to the $\gamma$ vibrational ones.
However, it is interesting to observe that they in fact encountered 
the avoided crossing with a $\gamma$-vibrational mode, 
similar to the one shown in Fig.~2(f),
and obtained a collective path that turns into the triaxial plane
in their calculation for states with angular momentum $I=2$.

\section{Concluding Remarks}

We have applied the ASCC method to the oblate-prolate shape coexistence 
phenomena in  $^{68}$Se and $^{72}$Kr.
It was found that the self-consistent collective paths run approximately 
along the valley connecting the oblate and prolate local minima 
in the collective potential energy landscape. 
This is the first time that the self-consistent
collective paths between the oblate and prolate minima have been 
obtained for realistic situations 
starting from the microscopic P+Q Hamiltonian.
Recently, the generator coordinate method has often been used
to describe a variety of shape coexistence phenomena, with
$\beta$ employed as the generator coordinate\cite{dug03,ben04}. 
The triaxial shape vibrational degrees of freedom are also ignored
in the extensive variational calculations 
by T\"ubingen group\cite{pet99,pet02}.
The result of the ASCC calculation, however, strongly indicates 
the necessity of taking into account the $\gamma$ degree of freedom,
at least for the purpose of describing the oblate-prolate shape 
coexistence in $^{68}$Se and $^{72}$Kr. 
In order to evaluate the mixing effects between the oblate and prolate 
shapes taking into account the triaxial deformation dynamics, 
we have to quantize the classical collective Hamiltonian obtained in this paper and solve the resulting collective Schr\"odinger equation. 
This will be the subject of the succeeding paper\cite{kob05}.

\section*{Acknowledgements}

This work was done as a part of Japan-U.S. Cooperative Science Program
``Mean-Field Approach to Collective Excitations in Unstable
Medium-Mass and Heavy Nuclei," 
and supported by 
the Grant-in-Aid for the 21st Century COE ``Center for Diversity and
Universality in Physics" from the Ministry of Education, Culture, Sports,
Science and Technology (MEXT) of Japan
and also 
by the Grant-in-Aid  for Scientific Research (Nos.~14540250 and 14740146) 
from the Japan Society for the Promotion of Science.
The numerical calculations were performed on the NEC SX-5 supercomputer
at Yukawa Institute for Theoretical Physics, Kyoto University.

\appendix

\section{Explicit expressions of the quasiparticle matrix elements}

Combining the successive Bogoliubov transformations,
(3.8) and (3.18), the quasiparticles,
$a_{\mu}^{\dag}(q)$ and $a_{\mu}(q)$, associated with
the state $|\phi(q)\rangle$,
can be written in terms of the nucleon operator,
$d_{k}^{\dag}$ and $d_{\bar k}$, as
\begin{equation}
\left(
 \begin{array}{c}
   a_{\mu}^{\dag} (q)\\
   a_{\nmu}(q)
 \end{array}
\right)
 = \sum_{k}
\left(
 \begin{array}{cc}
   U_{\mu k}(q)&V_{\mu \nk}(q)\\
   V_{\nmu k}(q)&U_{\nmu \nk}(q) 
 \end{array}
\right)
\left(
\begin{array}{c}
   d_{k}^{\dag} \\
   d_{\nk}
 \end{array}
\right)
\end{equation}
Making use of the inverse transformation,
\begin{equation}
\left(
 \begin{array}{c}
   d_{k}^{\dag}\\
   d_{\nk}
 \end{array}
\right)
 = \sum_{\mu}
\left(
 \begin{array}{cc}
   U_{k \mu}(q)&V_{k \nmu}(q)\\
   V_{\nk \mu}(q)&U_{\nk \nmu}(q) 
 \end{array}
\right)
\left(
\begin{array}{c}
   a_{\mu}^{\dag}(q) \\
   a_{\nmu}(q)
 \end{array}
\right),
\end{equation}
one can easily derive the explicit expressions of the expectation 
values and the matrix elements of the operators, ${\hat F}_{s}^{(\pm)}$, 
appearing in Eq.~(\ref{Fhatp}): 
\begin{eqnarray}
\bra\phiq {\hat F}_{s=1,2}^{(+)}\ket\phiq &=& 
- 2 \sum_{\mu}\sum_{k} \rbra{k{\bar k}} A_{\tau=n,p}^{(+)}\rket{0} 
U_{k \mu}(q)V_{\nk \mu}(q),
\nonumber \\
F_{s=1,2}^{(\pm)}(\mu \nnu) &=& 
\sum_{k} \rbra{k{\bar k}} A_{\tau=n,p}^{(\pm)}\rket{0}
\Big( U_{k \mu}(q)U_{\nk \nnu}(q)\pm V_{k \nnu}(q)V_{\nk \mu}(q)\Big), 
\nonumber \\
F_{B,s=1,2}^{(\pm)}(\mu \nu) &=&
\sum_{k} \rbra{k{\bar k}} A_{\tau=n,p}^{(\pm)}\rket{0} 
\Big( U_{k \mu }(q)V_{\nk \nu}(q)\pm U_{k\nu}(q)V_{\nk \mu}(q)\Big),
\nonumber \\
\bra\phiq {\hat F}_{s=3,5}^{(+)}\ket\phiq &=& 
2 \sum_{\nmu}\sum_{kl} \rbra{k} D_{2,K=0,2}^{(+)}\rket{l}
V_{k \nmu}(q)V_{l \nmu}(q),\nonumber \\
F_{s=3,5}^{(+)}(\mu \nnu) &=& 
\sum_{kl} \rbra{k} D_{2,K=0,2}^{(+)}\rket{l}
\Big(U_{k \mu }(q)V_{l \nnu }(q)+U_{ k \nu}(q)V_{l \nmu}(q)\Big), 
\nonumber \\
F_{B,s=3,5}^{(+)}(\mu \nu) &=&
\sum_{kl} \rbra{k} D_{2,K=0,2}^{(+)}\rket{l}
\Big(U_{k \mu }(q)U_{l \nu }(q) - V_{k \nnu }(q)V_{l \nmu }(q)\Big), 
\end{eqnarray}
The expectation values of the anti-Hermitian operators,
$A_{\tau=n,p}^{(-)}$, vanish.
The quantities, 
$\rbra{k} D_{2K}^{(+)}\rket{l}$, etc., appearing in the above expressions
denote matrix elements between the
single-particle states defined by Eq.~(\ref{sigbasis}):
\begin{eqnarray}
\rbra{k} D_{2K}^{(+)}\rket{l} &\equiv&
\rbra{0}d_k D_{2K}^{(+)}d_l^{\dag}\rket{0}, \nonumber \\
\rbra{k \nk} A_{\tau}^{(\pm)}\rket{0} &\equiv&
\rbra{0}d_{\nk}d_{k} A_{\tau}^{(\pm)}\rket{0},~~~{\rm etc}.,
\end{eqnarray}
where $\rket{0}$ is the vacuum for the nucleon operators $(d^{\dag}, d)$.
The matrix elements of the Bogoliubov transformations, (A.1) and (A.2),
possess the following symmetries:
\begin{equation}
U_{\mu k}=U_{\nmu \nk}=U_{k \mu}=U_{\nk \nmu},~~
V_{\mu \nk}=-V_{\nmu k}=V_{k \nmu}=-V_{\nk \mu}.
\end{equation}
It is also easily seen that equalities,
\begin{equation}
\rbra{\nk} D_{2,K=0,2}^{(+)}\rket{\nl} =
\rbra{k} D_{2,K=0,2}^{(+)}\rket{l},
\end{equation}
\begin{equation}
~~~ F_{B,s}^{(\pm)}(\nmu \nnu) = F_{B,s}^{(\pm)}(\mu \nu),
~~~ F_{B,s}^{(\pm)}(\nu \mu)= \pm F_{B,s}^{(\pm)}(\mu \nu),
\end{equation}
hold for the pairing and quadrupole operators,
$\{A_{n}^{(\pm)},A_{p}^{(\pm)},D_{20}^{(+)},D_{22}^{(+)}\}$,
under consideration.
The expectation values, 
$\bra\phiq {\hat N}_{\tau}\ket\phiq$,
and the matrix elements, 
$N_{\tau}(\mu)$ and $N_{B,\tau}(\mu)$, of the 
neutron and proton number operators are readily obtained 
from those of ${\hat F}_{s=3}^{(+)}$
by replacing   $\rbra{k} D_{20}^{(+)}\rket{l}$
with $\delta_{kl}$,
and restricting the sum over the single-particle index $k$
to neutrons or protons.

\section{Solving the moving frame HB equation}

We solve the moving frame HB equation using a method similar
to the imaginary time method\cite{dav80}.
Let $\ket{\phi^{(i)}(q)}$ be the state vector in the iterative step $i$.
We first calculate the mean-field Hamiltonian associate with it:
\begin{eqnarray}
{\hat h}^{(i)}(q) &=& 
   \sum_k \epsilon_k (d_k^{\dag}d_k + d_{\bar k}^{\dag}d_{\bar k}) 
  -\sum_{s}\kappa_{s}\expect{ {\hat F}_{s}^{(+)}}^{(i)}{\hat F}_{s}^{(+)}, 
\nonumber \\
\expect{ {\hat F}_{s}^{(+)}}^{(i)} &\equiv& 
\bra{\phi^{(i)}(q)}{\hat F}_{s}^{(+)}\ket{\phi^{(i)}(q)}.
\end{eqnarray}
Using the quasiparticle operators
$b_{\mu}^{(i)\dag}$ and $b_{\mu}^{(i)}$ defined by
\begin{equation}
b_{\mu}^{(i)}\ket{\phi^{(i)}(q)} = 0, 
\end{equation}
we then generate a state vector in the $(i+1)$ step as
\begin{eqnarray}
\ket{{\phi^{(i+1)}(q)}} & \equiv & \exp{\hat{X}^{(i+1)}}\ket{{\phi^{(i)}(q)} } 
\nonumber \\
{\hat X^{(i+1)}} = 
&-&\varepsilon \Big({\hat h}^{(i)}(q)-\sum_{\tau}\lambda_{\tau}^{(i+1)}(q)
{\hat N}_{\tau}-\mu^{(i+1)}(q){\hat Q}(q) \Big)_{+} \nonumber \\
&+&\varepsilon \Big({\hat h}^{(i)}(q)-\sum_{\tau}\lambda_{\tau}^{(i+1)}(q)
{\hat N}_{\tau}-\mu^{(i+1)}(q){\hat Q}(q) \Big)_{-} \nonumber \\
&\equiv& \sum_{\mu \nnu} x_{\mu \nnu}^{(i+1)}\Big({\xmndag -\xmn }\Big), 
\end{eqnarray}
where $\varepsilon$ is a small parameter,
\begin{equation}
\xmndag = b_{\mu}^{(i)\dag}b_{\nnu}^{(i)\dag},~~~
\xmn = b_{\nnu}^{(i)}b_{\mu}^{(i)},
\end{equation}
and the subscripts, $+$ and $-$, 
denote the two-quasiparticle creation and annihilation parts 
of the operator in the parenthesis, respectively.
It should be noted that, in contrast to the conventinal imaginary time method,
the unitary operator, $\exp {\hat X^{(i+1)}}$, is used here
so that the normalization is preserved during the iteration.
The Lagrange multipliers  $\lambda_{\tau}^{(i+1)}(q)$ and $\mu^{(i+1)}(q)$ are
determined by the constraint equations,  

\begin{eqnarray}
\bra{{\phi^{(i+1)}(q)}}{\hat N}_{\tau}{\ket {\phi^{(i+1)}(q)}}&=&N_{\tau}^{(0)} \nonumber \\
\bra{{\phi^{(i+1)}(q)}}{\hat Q}(q-\delta q)
{\ket {\phi^{(i+1)}(q)}} &=& \delta q,
\label{chbcon1}
\end{eqnarray}
where $N_n^{(0)}$ and $N_p^{(0)}$ are the neutron and proton numbers of the
nucleus under consideration.
Similar but slightly different constraints are utilized 
by Almehed and Walet\cite{alm04a}.
Expanding the left-hand sides up to first order in $x^{(i+1)}$, 
we obtain equations determining them: 
\begin{eqnarray}
\left(
    \begin{array}{ccc}
       (N_{n},N_{n}) &(N_{n},N_{p}) &(N_{n},Q(q)) \\
       (N_{p},N_{n}) &(N_{p},N_{p}) &(N_{p},Q(q)) \\
       (Q(q-\delta q),N_{n}) &(Q(q-\delta q),N_{p}) &(Q(q-\delta q),Q(q))
    \end{array}
\right)
\left(
    \begin{array}{c}
     \lambda_{n}^{(i+1)}(q) \\
     \lambda_{p}^{(i+1)}(q) \\
     \mu^{(i+1)}(q)
    \end{array}
\right) \nonumber \\
=
\left(
    \begin{array}{c}
     (N_{n}^{(0)}-\expect{{\hat N}_{n}}^{(i)})/2\varepsilon
     +(h^{(i)}(q),N_{n}) \\
     (N_{p}^{(0)}- \expect{{\hat N}_{p}}^{(i)})/2\varepsilon
     +(h^{(i)}(q),N_{p})\\
     (\delta q-\expect{{\hat Q}(q-\delta q)}^{(i)})/2\varepsilon
     + (h^{(i)}(q),Q(q-\delta q)),   
    \end{array}
\right),
\label{constraint}
\end{eqnarray}
where the quantities, 
$(N_{\tau},N_{\tau'}), (N_{\tau},Q(q))$ etc., 
are defined by (3.33), except that the coefficients, 
$N_{\tau}(\mu), Q_{\mu\nnu}(q)$ etc.,  
involved in these quantities are here defined with respect to 
the two-quasiparticle creation and annihilation operators, 
$\xmndag$ and $\xmn$.
Using the state vector $\ket{{\phi^{(i+1)}(q)}}$, 
we calculate the mean-field Hamiltonian ${\hat h}^{(i+1)}(q)$ in the
$(i+1)$ step, and repeat the above procedure until convergence is attained. 
The mean-field Hamiltonian thus obtained takes the following form: 
\begin{eqnarray}
{\hat h}_{M}(q) &=& {\hat h}(q)-\sum_{\tau}\lambda_{\tau}(q){\hat N}_{\tau}
-\mu(q){\hat Q}(q) \nonumber \\ 
&=& \bra\phiq \hat{h}_{M}(q) \ket\phiq
 + \sum_{\mu \nu}h_{\mu \nu}(q)
\Big( b_{\mu}^{\dag}(q)b_{\nu}(q) + b_{\nmu}^{\dag}(q)b_{\nnu}(q)\Big).
\label{qph}
\end{eqnarray}
Finally we introduce the quasiparticle operators, 
$a^{\dag}_{\mu}(q)$ and $a_{\mu}(q)$, that diagonalize ${\hat h}_{M}(q)$: 
\begin{equation}
{\hat h}_{M}(q) = \bra\phiq \hat{h}_{M}(q) \ket\phiq
+ \sum_{\mu} E_{\mu}(q)
\Big(a_{\mu}^{\dag}(q)a_{\mu}(q) + a_{\nmu}^{\dag}(q)a_{\nmu}(q)\Big).
\end{equation}
It is easy to see that $\mu(q)=\partial V/\partial q$.
In actual calculations, the above procedure is a part of the double iterative
algorithm described in section 4. Namely, we carry out the above iterative 
procedure using the constraint operator ${\hat Q}(q)^{(n)}$ 
that is obtained in the $n$-th iteration step determining the infinitesimal 
generators, ${\hat Q}(p)$ and ${\hat P}(q)$.

\begin{figure}
\centerline{\includegraphics[width=14 cm, height=19 cm]{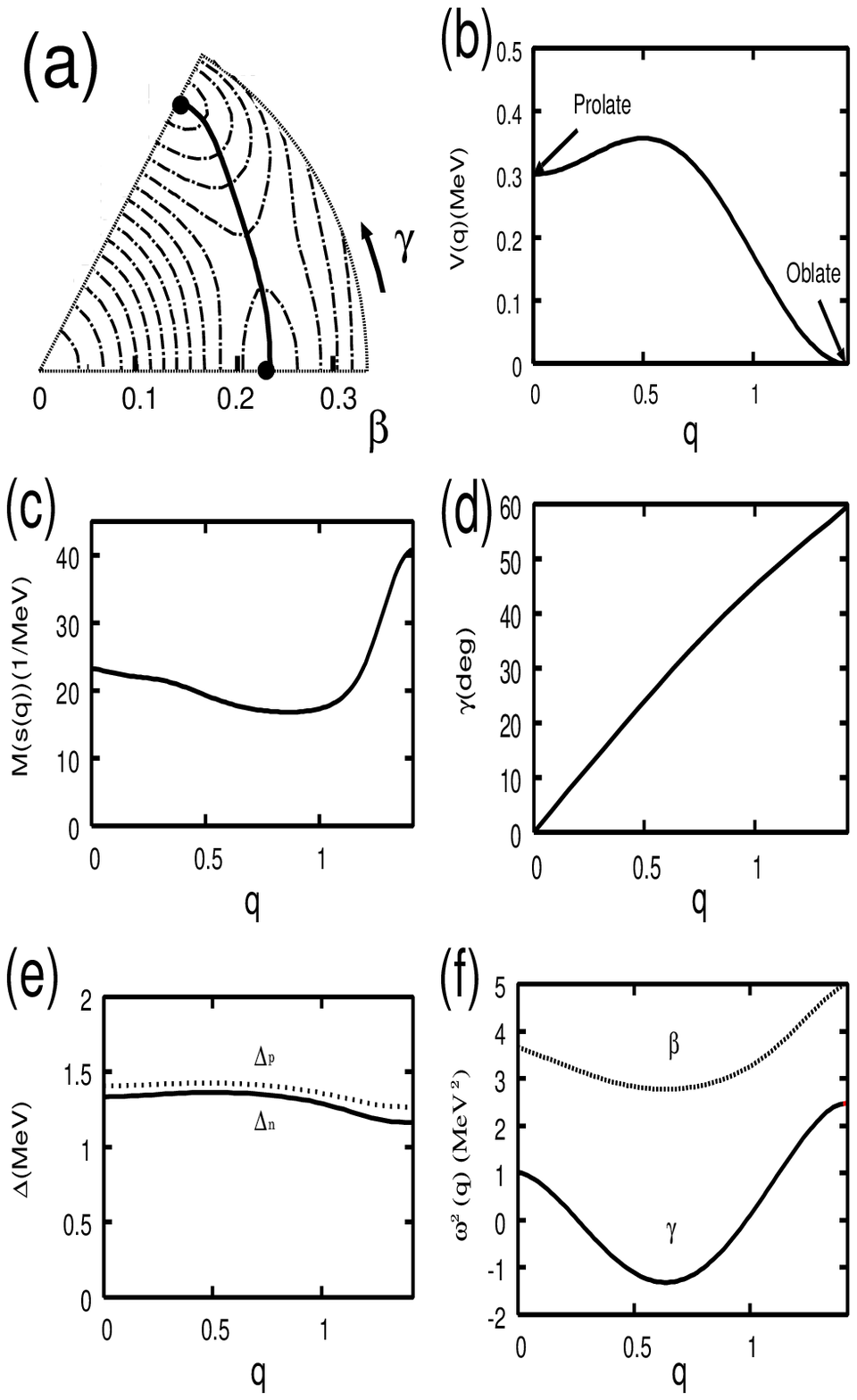}}
\caption{
Results of calculation for $^{68}$Se.
(a)~The bold curve represents the ASCC path 
projected onto the $(\beta,\gamma)$ plane,
which connects the oblate and the prolate minima 
designated by filled circles. 
The contour lines are calculated using the conventional
constrained HB method and plotted for every 50 keV. 
(b)~Collective potential $V(q)$
plotted as a function of the collective coordinate $q$.
Here the origin of $q$ is chosen to coincide with the prolate local minimum, 
and its scale is defined such that the collective mass is given by $M(q)=1$.
(c)~Collective mass $M(s(q))$ with respect to the
geometrical length $s(q)$ along the collective path 
in the $(\beta, \gamma)$ plane, plotted as a function of $q$.
(d)~The triaxiality parameter $\gamma$ as a function of $q$.
(e)~Neutron and proton pairing gaps, $\Delta_n$ and $\Delta_p$, 
as functions of $q$.
(f)~Lowest two eigen-frequencies squared (i.e., $\omega^2=BC$)
of the moving frame RPA, plotted as functions of $q$.
These modes at triaxial deformed shapes are more general than the
ordinary $\beta$- and $\gamma$-vibrations at the oblate and the prolate 
limits, and contain both components. 
The symbols, $\beta$ and $\gamma$, are used, however, 
in order to indicate the major components of the moving frame RPA modes.
}
\end{figure}

\begin{figure}
\centerline{\includegraphics[width=14 cm, height=19 cm]{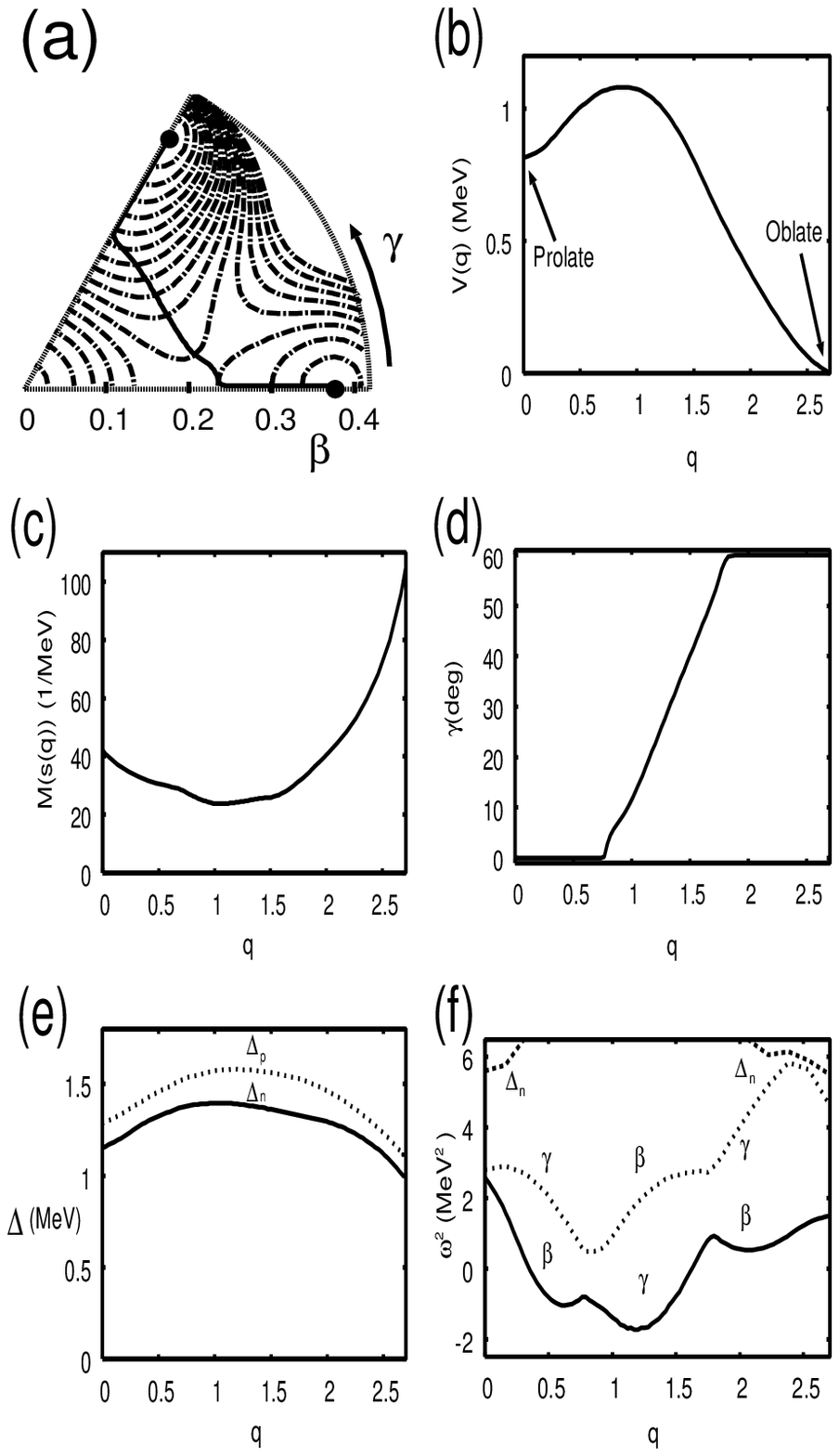}}
\caption{
Results of calculation for $^{72}$Kr.
Notations are the same as in Fig.~1 except for the followings:
In (a), the contour lines are plotted for every 100 keV.
In (f), the lowest three eigen-frequencies squared (i.e., $\omega^2=BC$)
of the moving frame RPA, are plotted as functions of $q$.
As mentioned in the caption to Fig.~1, 
these modes at triaxial deformed shapes are more general than the
ordinary $\beta$- and $\gamma$-vibrations at the oblate and the prolate 
limits, and contain both components. 
The symbols, $\beta$ and $\gamma$, are used, however, 
in order to indicate the major components of the moving frame RPA modes.
Likewise, the symbols $\Delta_n$ is used to indicate that
the major component is the neutron pairing vibrational mode.
}
\end{figure}


\begin{figure}   
\begin{center}
    \begin{tabular}{cc}
\includegraphics[width=70mm,height=70mm]{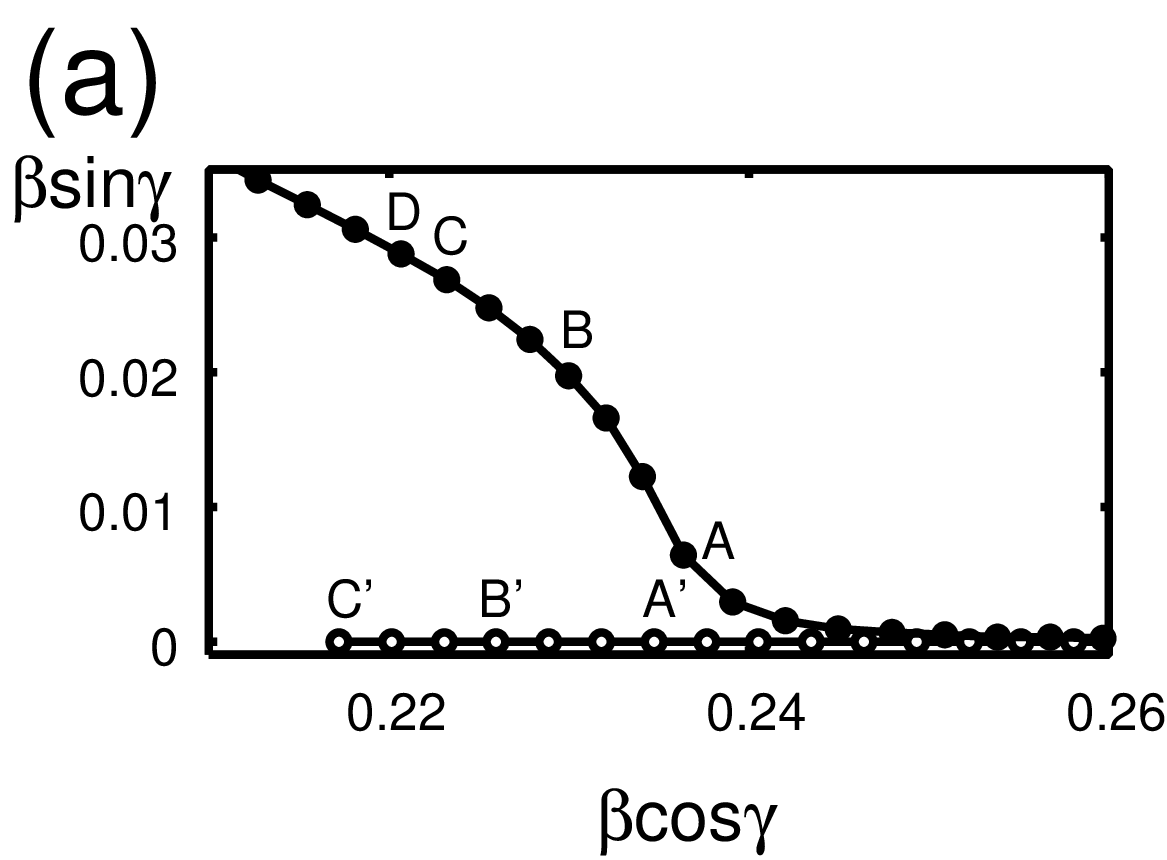} &
\includegraphics[width=70mm,height=70mm]{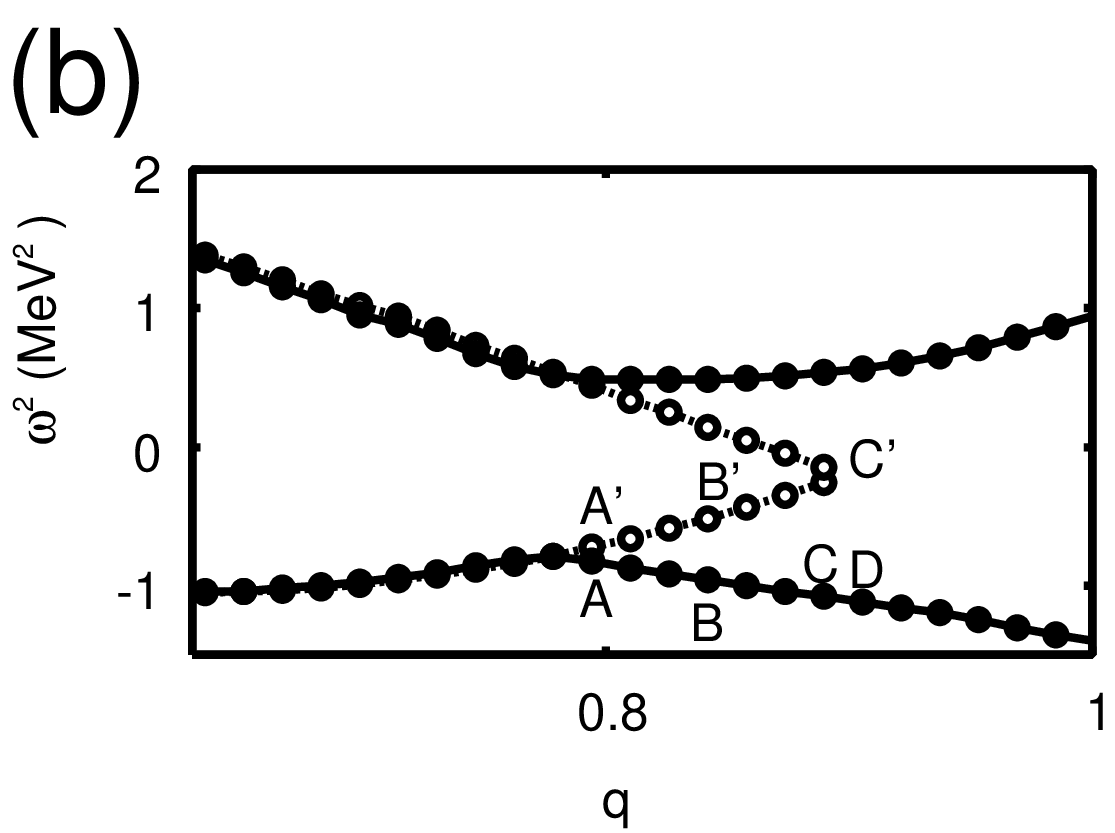} 
    \end{tabular}
\vspace{1cm}
\caption{
Enlargement of the turn over region of Figs.~2(a) and 2(f) for $^{72}$Kr, 
where the collective path turns into the $\gamma$ direction.
The step size $\delta q=0.0157$ and $\varepsilon=0.1$ are used
in the numerical calculation.  Every steps in $\delta q$ are 
represented by filled circles and connected by solid lines.
The points designated A, B, C, D on the collective path in (a) 
correspond to those in (b) which
displays the squared frequencies, $\omega^2$, of the lowest two solutions 
of the moving frame QRPA equations as 
functions of the collective coordinate $q$.
The open circles represent those obtained
in the calculation with $\varepsilon=0$ where the mixing effects between 
the $K$=$0$ and 2 components are totally ignored.
The points designated A', B', C' in the latter calculation 
correspond to the points A, B, C in the former calculation.
In the latter calculation, 
we cannot get the point corresponding to D because the problem
discussed in the text occurs in the numerical algorithm.
It was checked that the same collective path is
obtained with use of $\delta q=0.0314$ except that the distances between the
successive points are doubled.  
}
\end{center}
\end{figure}

\begin{figure}   
\centerline{\includegraphics[width=14 cm, height=10 cm]{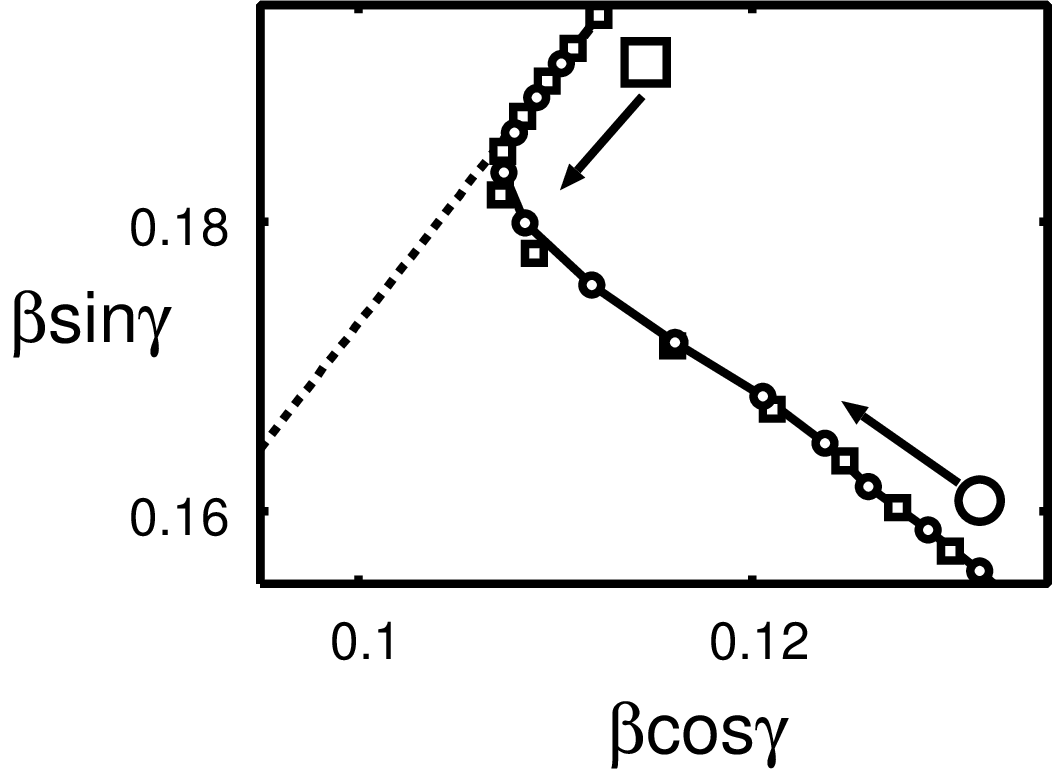}}
\caption{
Enlargement of the turn over region of Figs.~2(a) for $^{72}$Kr, 
where the collective path (solid curve) coming from the prolate minimum 
merges with the $\gamma=60^{\circ}$ axis (dotted line).
Every steps in $\delta q$ are represented by open circles and 
connected by solid lines.
For comparison, the result of calculation starting from the oblate minimum
and moving in the opposite direction is shown by open squares.
Slight deviations from the solid curve indicate the degree of precision of
the present numerical calculation.
The step size $\delta q=0.0157$ and $\varepsilon=0.1$ are used
in both cases. The collective path obtained by these different calculations 
well agree with each other.
}
\end{figure}


\begin{figure}   
\begin{center}
    \begin{tabular}{cc}
\includegraphics[width=70mm,height=70mm]{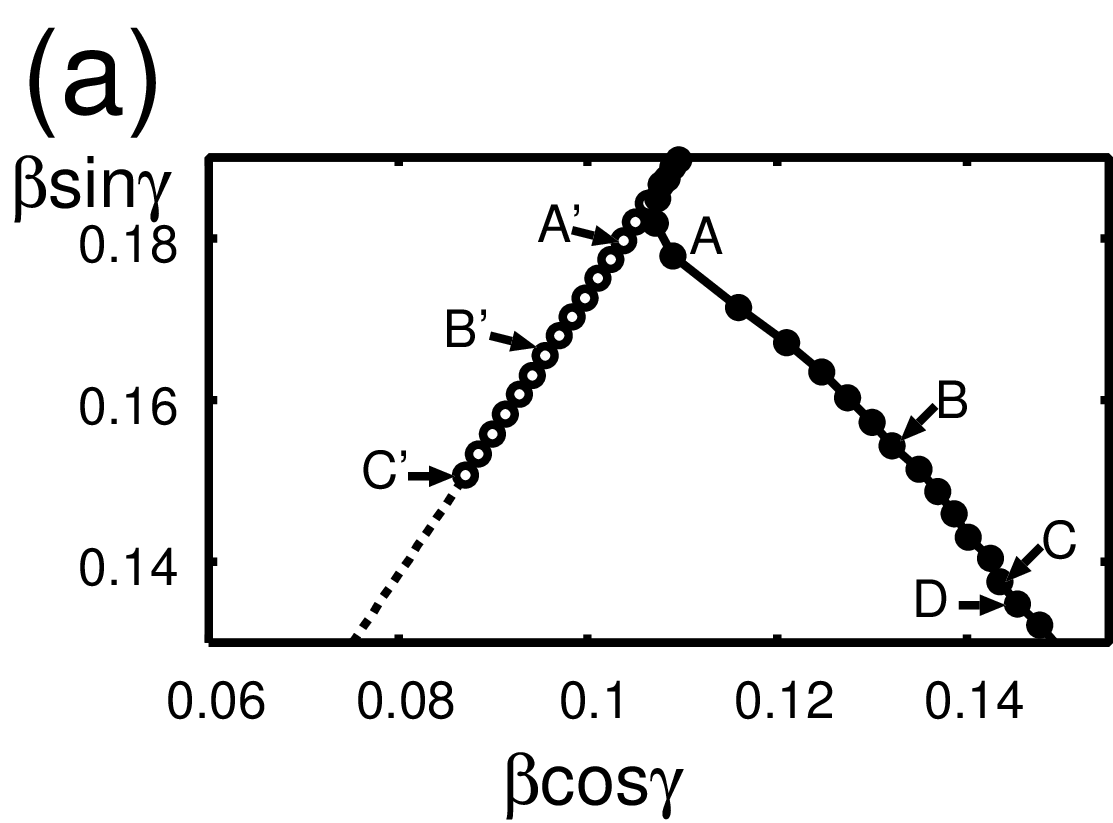} &
\includegraphics[width=70mm,height=70mm]{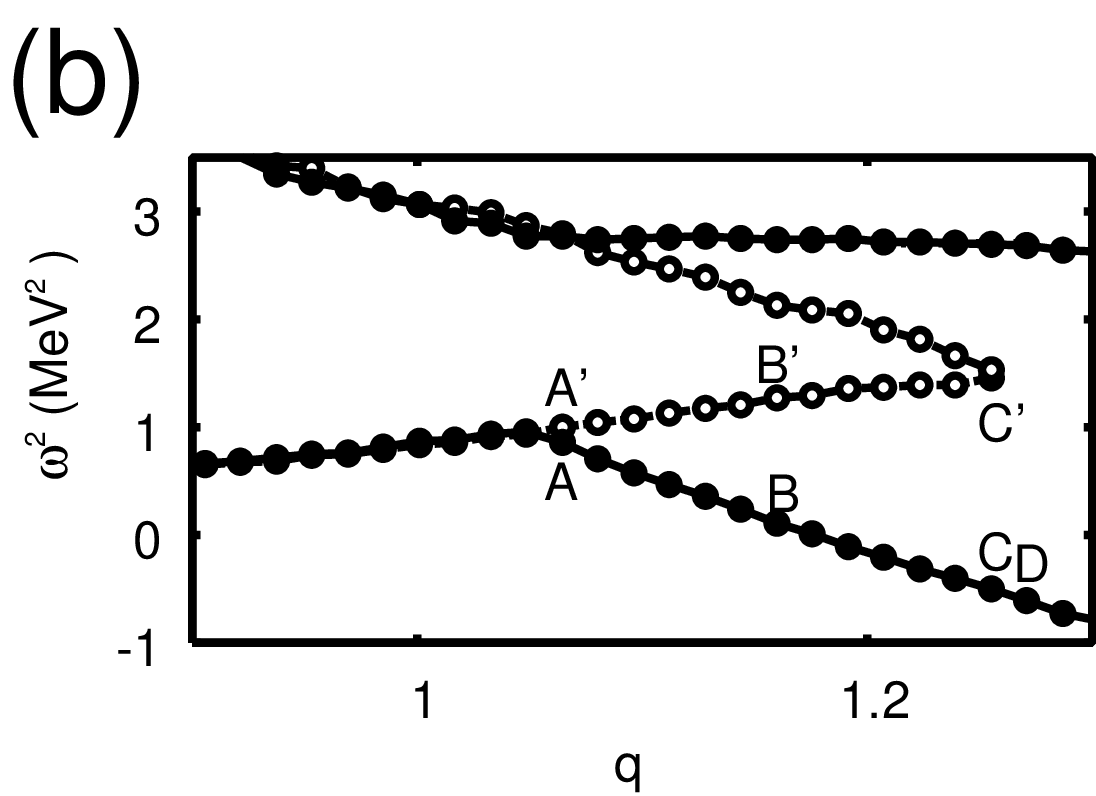} 
    \end{tabular}
\vspace{1cm}
\caption{
Enlargement of the turn over region of Figs.~2(a) and 2(f) for $^{72}$Kr, 
where the collective path (solid curve) coming from the oblate minimum 
starts to deviate from the $\gamma=60^{\circ}$ axis (dotted line).
The numerical calculation was done starting from the oblate minimum
and using the step size $\delta q=0.0157$ and $\varepsilon=0.1$.
Every steps in $\delta q$ are represented by filled circles and 
connected by solid lines.
The points designated A, B, C, D on the collective path in (a) 
correspond to those in (b) which
displays the squared frequencies, $\omega^2$, of the lowest two solutions 
of the moving frame QRPA equations as 
functions of the collective coordinate $q$.
Note that the values of $q$ in this figure are measured from the
oblate minimum.
The open circles represent those obtained
in the calculation with $\varepsilon=0$ where the mixing effects between 
the $K$=$0$ and 2 components are totally ignored.
The points designated A', B', C' in the latter calculation 
correspond to the points A, B, C in the former calculation.
In the latter calculation, 
we cannot get the point corresponding to D because the problem
discussed in the text occurs in the numerical algorithm.
Slight wiggles along the successive points seen in (b) are 
due to numerical error, and they indicate the degree of precision of
the present numerical calculation.
It was checked that the same collective path is
obtained with use of $\delta q=0.0314$ except that the distances between the
successive points are doubled.  
}
\end{center}
\end{figure}

\end{document}